\documentclass[printer]{aa}
\usepackage{amssymb}
\usepackage{amsmath}
\usepackage{natbib}
\bibpunct{(}{)}{;}{a}{}{,} 
\usepackage{graphicx}
\usepackage{txfonts}
\usepackage{color}
\usepackage{epstopdf}
\usepackage{wasysym}
\usepackage[utf8]{inputenc}
\DeclareUnicodeCharacter{2032}{-}
\usepackage{color,hyperref}
\definecolor{darkblue}{rgb}{0.,0.,1.0}
\hypersetup{colorlinks,breaklinks,
           linkcolor=darkblue,urlcolor=darkblue,
           anchorcolor=darkblue,citecolor=darkblue}

\usepackage{ulem}

\begin{document}
\title{Constraining the accretion flow density profile near Sgr~A* using the $L'$-band emission of the S2 star}

\author{S. Elaheh Hosseini\inst{1,2}, Michal Zaja\v{c}ek\inst{3, 2, 1}, Andreas Eckart\inst{1,2}, Nadeen B. Sabha\inst{4, 1}, and Lucas Labadie\inst{1}}

\institute{I. Physikalisches Institut der Universit\"at zu K\"oln, Z\"ulpicher Strasse 77, D-50937 K\"oln, Germany \and Max-Planck-Institut f\"ur Radioastronomie (MPIfR), Auf dem H\"ugel 69, D-53121 Bonn, Germany \and Centre for Theoretical Physics, Polish Academy of Sciences, Al. Lotnikow 32/46, 02-668 Warsaw, Poland \and Institut für Astro- und Teilchenphysik, Universität Innsbruck, Technikerstrasse 25/8, A-6929 Innsbruck, Austria}

\authorrunning{S. E. Hosseini et al.}

\titlerunning{Constraining the accretion flow density profile near Sgr~A*}

\date{Received 21.03.2020; accepted 28.09.2020}

\abstract{ The density of the ambient medium around a supermassive black hole, and the way it varies with distance, plays an important role in understanding the inflow-outflow mechanisms in the Galactic Centre. This dependence is often fitted by spherical power-law profiles based on observations in the X-ray, infrared, submm and radio domains.}{ Nevertheless, the density profile is poorly constrained at the intermediate scales of 1000 Schwarzschild radii. Here we independently constrain the spherical density profile using the stellar bow shock of the star S2 which orbits the supermassive black hole at the Galactic Centre with the pericentre distance of  $14.4$ mas ($\sim 1500$ R\textsubscript{s} ).}{ Assuming an elliptical orbit, we apply analytical relations from celestial mechanics and the theory of bow shocks that are at ram pressure equilibrium. We analyse the measured infrared flux density and magnitudes of S2 in the $L'$-band (3.8 micron) obtained over seven epochs in the years between 2004-2018. We use this to put an upper limit on the emission from S2’s associated putative bow shock and to constrain the density profile of the ambient medium.}{ We detect no significant change in S2 flux density between 2004 and the recent periapse in May 2018. The intrinsic flux variability of S2 is at the level of 2-3\% only. Based on the dust-extinction model, the upper limit on the number density at the S2 periapse is $\sim 1.87\times 10^9\,{\rm cm^{-3}}$, which yields the density slope of at most $3.20$. Using the synchrotron bow-shock emission, we obtain the ambient density of $\lesssim 1.01 \times 10^5\,{\rm cm^{-3}}$ and the slope of $\lesssim 1.47$. These values are consistent with a wide variety of media from hot accretion flows to potentially colder and denser medium comparable in properties to broad-line region clouds. Standard thin disc can be, however, excluded at the distance of S2 pericentre.}{ With the current photometry sensitivity of 0.01 mag, we are not able to make stringent constraints on the density of the ambient medium in the Galactic Centre using S2-star observations. Now, only the distinction between hot accretion flows and thin, cold discs is possible, where the latter can be excluded at the scale of the S2 periapse. Future observations of stars in the S cluster using instruments such as METIS@ELT with the photometric sensitivity of as much as $10^{-3}$ mag will allow to probe the Galactic Centre medium at intermediate scales down to as small densities as $\sim 700\,{\rm cm^{-3}}$ for the non-thermal bow-shock emission. The new instrumentation, in combination with discoveries of stars with smaller pericentre distances, will help to independently constrain the density profile around Sgr~A*.}


\maketitle
\section{Introduction}

 The Galactic Centre is a unique astrophysical setting where one can study the dynamical effects in one of the densest stellar clusters \citep{2005PhR...419...65A,2013degn.book.....M,2014CQGra..31x4007S} and the mutual interaction of stars, gaseous-dusty structures, the magnetic field, and the supermassive black hole (hereafter SMBH) \citep{2010RvMP...82.3121G,2017FoPh...47..553E}. The dynamical centre traced by orbiting stars coincides with the non-thermal compact and variable radio source Sgr~A* within $0.03''$\citep{1997ApJ...475L.111M}. Because of its well-constrained mass of $M_{BH} = (4.15 \pm 0.13 \pm 0.57)\times 10^6\,M_{\odot}$ \citep{Parsa2017} based mainly on monitoring of fast-moving S stars (see also \citealt{Schoedel2002,Ghez2008,Gillessen2009,2016ApJ...830...17B,2017ApJ...837...30G}), Sgr~A* is associated with the SMBH, leaving only a little space for alternative scenarios of its compact nature \citep{2017FoPh...47..553E}. Given its distance of $8.1\,{\rm kpc}$ \citep{Eisenhauer2003,reid2003,GravColl2019}, it is the nearest SMBH, which allows to perform high-precision astrometric and interferometric observations of Sgr~A* and its immediate surroundings. In recent years, this allowed to confirm general relativistic predictions by measuring the gravitational redshift \citep{2018AA...615L..15G,2019Sci...365..664D} as well as the Schwarzschild precession \citep{2020AA...636L...5G} by monitoring the bright S2 star (also referred to as S0-2) that orbits Sgr~A* with a period of 16.2 years.

Sgr~A* belongs to extremely low-luminous sources, with its bolometric luminosity eight to nine orders of magnitude below its Eddington limit \citep{1998ApJ...492..554N,2014ARAA..52..529Y}. Its broadband spectral energy distribution is best described by a class of radiatively inefficient accretion flows \citep[RIAFs; ][]{2003ApJ...598..301Y}. To better comprehend its low-luminous, diluted accretion flow, it is of a prime importance to constrain its density at large, intermediate, and eventually small spatial scales. This has been done by analysing radiative properties of its surroundings. At larger scales, the analysis of the X-ray bremsstrahlung emissivity profile of $1.3\,{\rm keV}$ plasma in the central hot bubble yielded the number density of $n_{\rm B}\approx 26\,\eta_{\rm f}^{-1/2}{\rm cm^{-3}}$ (with $\eta_{\rm f}$ being the filling factor) close to the Bondi radius \citep{2003ApJ...591..891B,2013Sci...341..981W}, which lies at $r_{\rm B}\approx 4''(T_{\rm a}/10^7\,{\rm K})^{-1}\approx 0.16\,{\rm pc}$ given the plasma temperature $T_{\rm a}$ \citep{2013Sci...341..981W}. The density constraints at the scale of 10--100 Schwarzschild radii (hereafter denoted as $r_{\rm s}$) were obtained by analysing the polarised millimeter and submillimeter emission, which implies non-thermal self-absorbed synchrotron radiation in the millimeter domain that becomes optically thin at submillimeter wavelengths  \citep{2000ApJ...534L.173A,2000ApJ...538L.121A,2003ApJ...588..331B}. \citet{2006ApJ...640..308M,2007ApJ...654L..57M} detected the Faraday rotation of the polarised submillimeter emission, based on which they constrain the accretion-rate of Sgr~A* to $\dot{M}\sim 2\times 10^{-9}-2\times 10^{-7}\,M_{\odot}\,{\rm yr^{-1}}$ that corresponds to the maximum number density of $n_{\rm a}^{\rm max}<1.4\times 10^7\,{\rm cm^{-3}}$ close to the event horizon. Generally, the hot accretion flow in the Galactic Centre is described by a power-law density profile $n\propto r^{-\gamma}$, where the power-law slope is typically inferred to be $\gamma\sim 0.5-1.0$ \citep{2013Sci...341..981W} in agreement with the RIAF-type flows. Spherical Bondi-type flow with $\gamma\sim 3/2$ is also consistent with the X-ray surface brightness profile up to 3'' from Sgr~A* \citep{2015AA...581A..64R}. Hence, the number densities of the ambient flow and the corresponding slopes are still quite uncertain, especially at intermediate spatial scales where the fast-moving S stars are located. This uncertainty was further enhanced by the possible detection of a cooler ($\sim 10^4\,{\rm K}$) and a denser disc with the number density of $\sim 10^5-10^6\,{\rm cm^{-3}}$ at the radius of $\sim 0.004\,{\rm pc}$ comparable to the semi-major axis of the bright S2 star \citep{2019Natur.570...83M}. However, the detection by \citet{2019Natur.570...83M} is questionable, especially the association of the recombination double-peak line of H30$\alpha$ at 1.3 mm with the gaseous material at milliparsec spatial scales. It is especially controversial since near-infrared line maps, e.g., of recombination Br$\gamma$ line that can trace ionised material of $10^4\,{\rm K}$, do not seem to imply any existence of such a compact disc-like structure. On the other hand, there are indications of denser material traced by Br$\gamma$ line \citep{2020ApJ...897...28P} and the blueshifted H30$\alpha$ line \citep{2019ApJ...872....2R} at larger spatial scales closer to the Bondi radius, which the study of \citet{2019Natur.570...83M} could be associated with. This underlines the need for better and independent constraints on the accretion-flow density at the intermediate range of radii.

The fast-moving stars of the S cluster could in principle be used to constrain the accretion flow density at intermediate scales. As fast-moving probes, they drive shocks into the ambient medium and provide radiative energy that can temporarily affect the radiative properties of the accretion flow as well as form localized density and extinction enhancements. Any detected flares linked to an orbiting star would thus provide a way to infer density and other properties of the intercepted ambient medium. \citet{2005ApJ...635L..45Q} analysed the possibility that the stellar wind shocks in the central arcsecond could contribute to the ambient TeV and nonthermal broadband emission. It was predicted that the interaction of the fast-moving B-type S2 star with the ambient hot accretion flow is expected to form a bow shock, which should result in a month-long thermal X-ray bremsstrahlung flare with the peak luminosity of $L_{\rm X}=4\times 10^{33}\,{\rm erg\,s^{-1}}$ \citep{Giannios+2013,2016MNRAS.459.2420C}. The contribution to the emerging X-ray flux is also expected to come from UV and optical photons of S2 that are Compton-upscattered by electrons of the inner hot RIAF \citep{2005AA...429L..33N}. According to these models, the S2 bow-shock X-ray luminosity is in principle comparable to the quiescent X-ray emission of Sgr~A* and hence could be detected with the current X-ray instruments. The non-detection of any significant X-ray flare around the pericentre crossings of S2 star in 2002 and 2018 is consistent with a diluted hot RIAF flow in the central arcsecond \citep{2003ApJ...598..301Y}. The low X-ray flux from the bow shock is also in agreement with numerical 3D adaptive mesh refinement simulations of the motion of S2 through the RIAF \citep{Schartmann+2018}.

The detection and the monitoring of the dusty G2 source or the Dusty S-cluster Object \citep[hereafter DSO; ][]{2012Natur.481...51G,2013AA...551A..18E,2014ApJ...796L...8W, Valencia_S__2015,peissker2020,2020Natur.577..337C} triggered an effort to predict and observe its potential interaction with the ambient environment. \citet{Narayan+2012} predicted the radio synchrotron emission from the DSO bow-shock to be comparable to the quiescent radio emission of Sgr~A*. In theory, electrons in the ambient accretion flow are expected to be accelerated in the bow-shock region. The main sources of uncertainty are the bow-shock size, which depends on the stellar-wind characteristics and the ambient density, and the magnetic field strength that is enhanced in the bow-shock. For the more realistic estimates of the DSO/G2  bow-shock size, assuming the low-mass star model, the radio synchrotron emission was calculated to be well below the radio emission of Sgr~A*, hence no detectable flare was expected \citep{Crumley+2013,2016MNRAS.455.1257Z}. This was observationally confirmed by the non-detection of any enhanced radio and millimeter emission in \citet{2015ApJ...802...69B} and \citet{2016MNRAS.458.2336B}, who constrain the DSO/G2 cross-section to $<2 \times 10^{29}\,{\rm cm^2}$. 

The bow-shock synchrotron emission can be generalized to the S stars, predicting their broad-band synchrotron spectrum with the peak close to 1 GHz and associated monochromatic light-curves \citep{Ginsburg+2016}. \citet{Ginsburg+2016} estimated that the synchrotron emission from S-star bow shocks can be comparable to the Sgr~A* radio (10 GHz) and infrared flux densities ($10^{14}$ GHz) for rather extreme combinations of the wind mass-loss rate and its terminal velocity -- $(\dot{m}_{\rm w}, v_{\rm w})=(10^{-5}\,M_{\odot}\,{\rm yr^{-1}},10^3\,{\rm km\,s^{-1}})$ and $(\dot{m}_{\rm w}, v_{\rm w})=(10^{-6}\,M_{\odot}\,{\rm yr^{-1}},4\times 10^3\,{\rm km\,s^{-1}})$. These values deviate from those inferred for S2 \citep{2008ApJ...672L.119M,2017ApJ...847..120H}: $(\dot{m}_{\rm w}, v_{\rm w})=(<3\times 10^{-7}\,M_{\odot}\,{\rm yr^{-1}}, 10^3\,{\rm km\,s^{-1}})$. For these observationally inferred values, the S star nonthermal flux densities are below those of Sgr~A*. Another, independent way to constrain the density of the ambient medium is the detection of deviation from the Keplerian ellipse due to hydrodynamic drag force. \citet{2019ApJ...871..126G} claimed to detect a significant deviation for DSO/G2 object due to the drag force, from which they inferred the density of $4\times 10^3\,{\rm cm^{-3}}$ at the DSO/G2 pericentre close to 1000$\,{\rm r_{\rm s}}$.

Here we report about the near-infrared $L'$-band emission of one of the brightest members of the S cluster - star S2. Close to its first observed pericentre passage in 2002, \citet{Clenet-2004} report a brightening of S2 in $L'$-band, which could be associated with its interaction with the ambient medium at the scale of 1500 Schwarzschild radii\footnote{The S2 pericentre distance according to \citet{2018AA...615L..15G} is $1513$ Schwarzschild radii.}. Based on our reanalysis of 2002 epoch and additional seven epochs, including close to the subsequent pericentre passage in 2018, we can exclude any significant brightening related to the pericentre passage of S2 in $L'$-band. The intrinsic fractional variability of the S2 lightcurve is only $\sim 2.5\%$. Using the model of the thermal dust emission in the shocked ambient medium, we place an upper limit on the ambient density of $\lesssim 10^9\,{\rm cm^{-3}}$, which can accommodate a wide variety of ambient media at the S2 pericentre distance of 1500 Schwarzschild radii, except for a standard thin cold disc. The bow-shock synchrotron model gives a tighter upper limit of $\sim 10^5\,{\rm cm^{-3}}$. An improved photometric sensitivity of the next-generation of instruments, in particular METIS@ELT, will help to further constrain the density at intermediate scales.

The paper is structured as follows. Section~\ref{Observation} describes the observations and data reduction methods used for this work. Then, we briefly discuss the photometry and its results for S2 in Section~\ref{Photometry} and Section~\ref{Photometry results}, respectively. In Section~\ref{observability} we constrain the slope and density of the ambient accretion flow using S2 star observations in $L'$ band. Subsection~\ref{extinction} is dedicated to the exploration of the extinction caused by the ambient gas and dust inside the Bondi radius.  In subsection~\ref{max_thermal_bowshock_contribution}, the upper limit of the ambient density is inferred based on the thermal bow-shock emission. Then in section~\ref{max_nonthermal_bowshock_contribution}, we constrain the ambient density based on the non-thermal bow-shock emission. We discuss the obtained density values in the broader context of the accretion flows and the Galactic Centre gaseous-dusty structures in section~\ref{Discussion}. Finally, in section~\ref{summary} we summarise the main results.  
\section{Observations and data reduction}
\label{Observation}  
Observations of the Galactic Centre in the near-infrared (NIR) were carried out with the European Southern Observatory (ESO) Very Large Telescope (VLT) UT1-YEPUN  on Paranal, Chile. The data was obtained from the infrared (IR) camera CONICA and the NAOS adaptive optics (AO) module, briefly NACO. The central parsec of the Galactic Centre has been frequently observed with the VLT in the $L'$-band (3.8 $\mu m$, 0.0271\arcsec /pixel). 

 In Table~\ref{S2-epochs}, we summarise the obtained data that will be further analysed in this study. The total time-coverage of the data set is from April 2004 till April 2018. The only applied criterion on the epoch selection is the imaging quality since a large number of the available data sets in the archive were observed during poor atmospheric conditions, in particular bad and fast seeing. We used the standard data reduction process that consists of the sky subtraction, flat-fielding and bad pixel correction for all imaging data and combined them via a shift-add algorithm to obtain the final array size of $28\arcsec\times28\arcsec $ for all epochs.

The first observed periapse of S2 occurred in 2002. The spatial resolution of NACO at 3.8 $\mu m$ is 0.096 \arcsec , while the distance between Sgr~A* and S2 in 2002 is approximately 0.040\arcsec, therefore it is impossible to obtain the emission of S2 from the images completely separated from Sgr~A* emission. Furthermore, we know that Sgr~A* is flaring in the observation which is done in 2002 \citep{Eckart_Nature}, and in 2002 and 2003 the NIR-excess extended source denoted as G1 is close enough to the S2-Sgr~A* system so that it cannot be disentangled from the Sgr~A*-S2 system \citep{Clenet-2004,2017ApJ...847...80W}.

\begin{figure*}[ht!]
    \centering
    \includegraphics[width=0.48\textwidth]{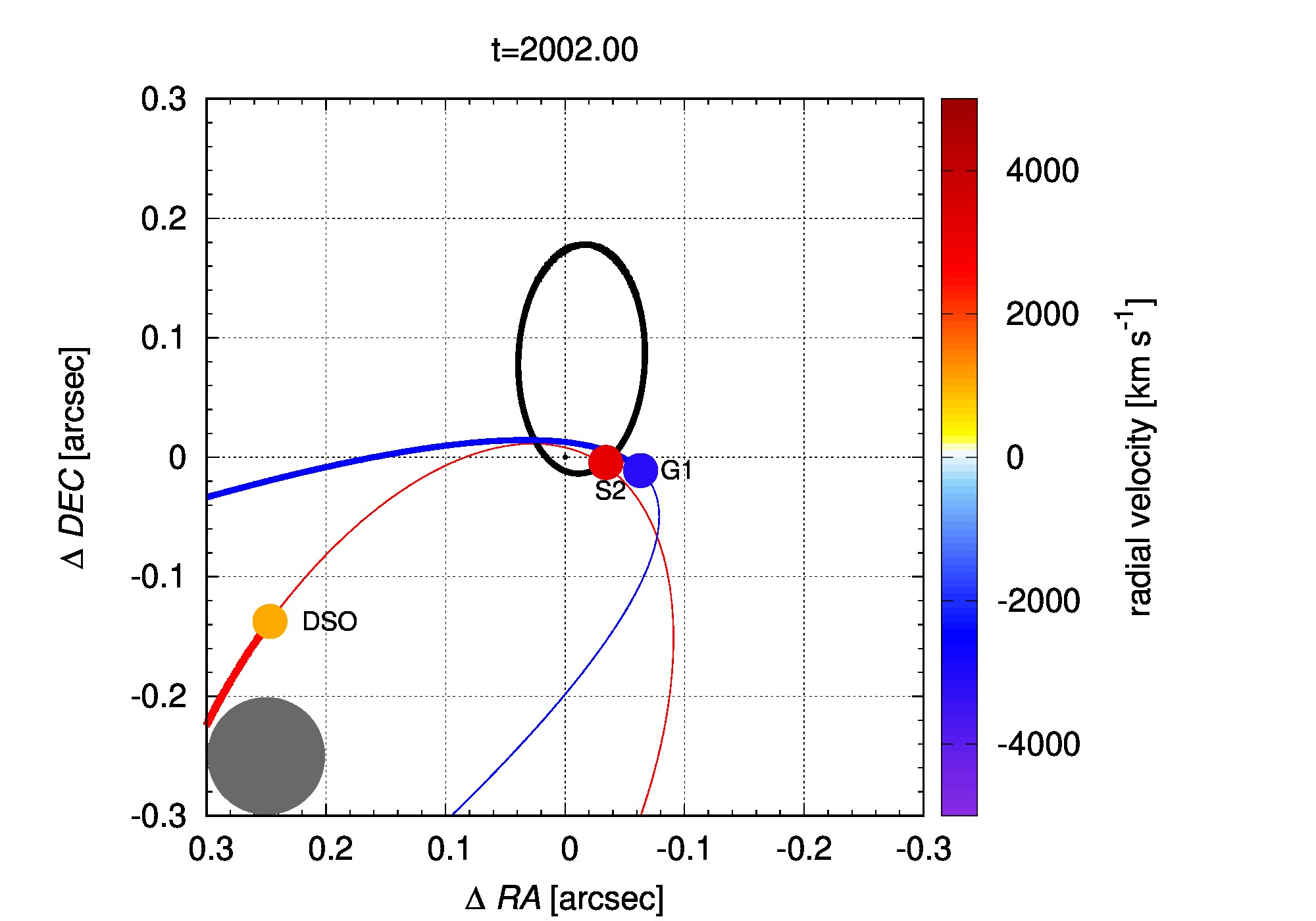}
    \includegraphics[width=0.48\textwidth]{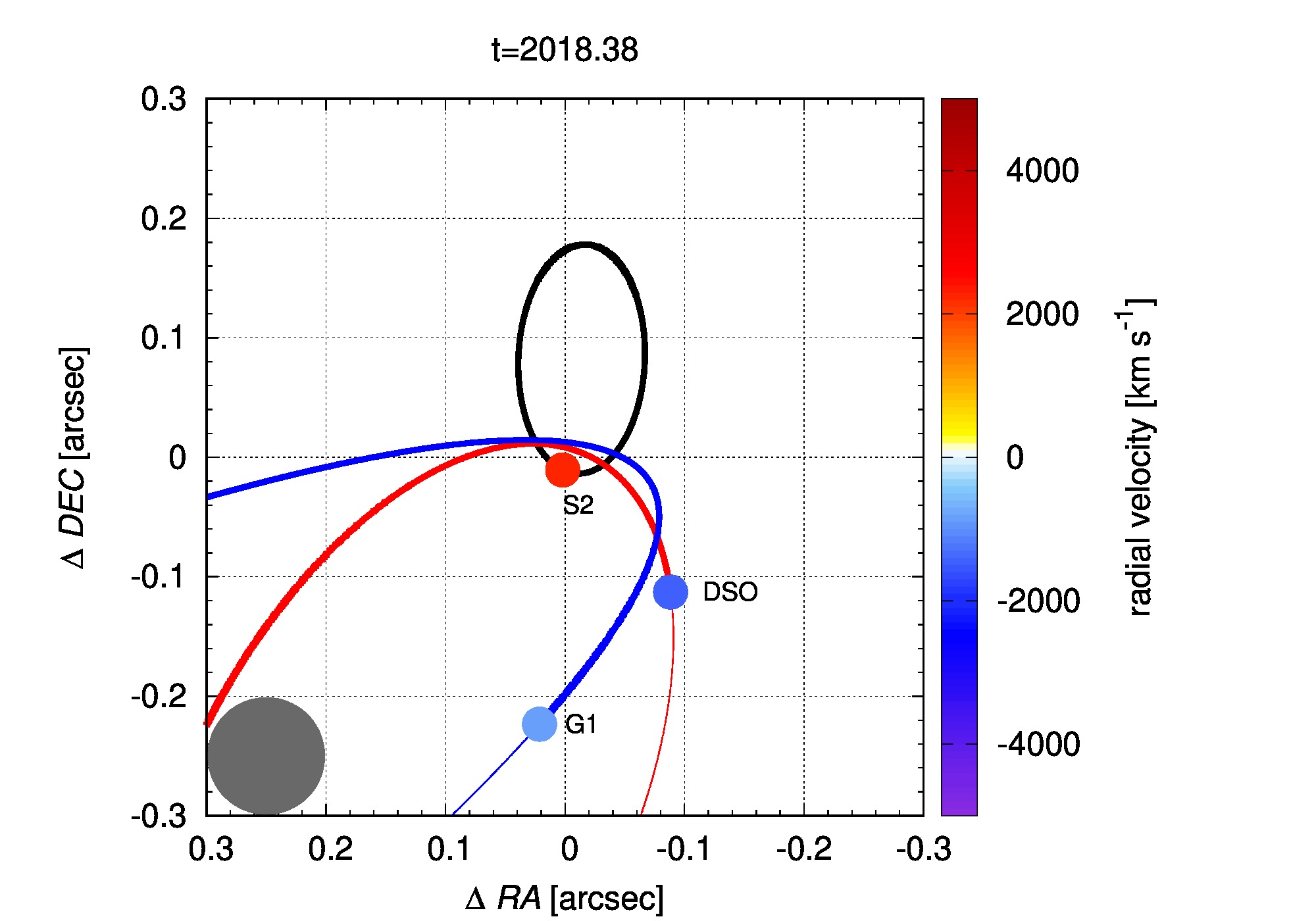}
    \caption{S2 star and known dusty objects close to its orbit. \textbf{Left panel:} The position of S2 and the dusty source G1 close to the pericentre of S2 around 2002. We also show the orbit of the DSO  (Dusty S-cluster Object, also known as G2). The points of the sources represent colour-coded radial velocities according to the axis on the right. The gray circle in the bottom left corner represents the diffraction limited FWHM of 98 mas corresponding to NACO $L'$-band. \textbf{Right panel:} The position of S2 and dusty source G1 at the pericentre passage of S2 at 2018.38. The orbital parameters were adopted from \citet{2018AA...615L..15G} (S2 star), \citet{2015ApJ...800..125V} (DSO source), and \citet{2017ApJ...847...80W} (G1 source).}
    \label{fig_orbits}
\end{figure*}

 We show the superposition of S2 and G1 in Fig.~\ref{fig_orbits}, where we plot the projected orbits of S2, G1, and DSO/G2. Right before the pericentre passage around $2002$, the mutual separation between S2 and G1 was less than $0.1$ arcsec, see Fig.~\ref{fig_orbits} (left panel). The G1 cloud is currently going to larger distances from Sgr~A* (more than $0.2$ arcsec, see Fig.~\ref{fig_orbits}, right panel), hence potential changes of the brightness of S2 after 2002 pericentre passage would be due to other effects, most likely due to bow-shock emission that we analyse in this contribution. Hence, to prevent further uncertainties in our analysis, we excluded the first observed periapse of S2.   

\subsection{Photometry}
\label{Photometry}

Flux densities of S2 and calibrators were determined via aperture photometry, using a circular aperture of 82 mas in $L'$-band. Dereddened fluxes have been computed with the extinction of $A_{L}=1.09\pm 0.13$ \citep{Fritz2011} and a zero magnitude value of $F_0(L) = 249$ Jy.

\begin{figure}[ht!]
\centering
\includegraphics[width=90mm]{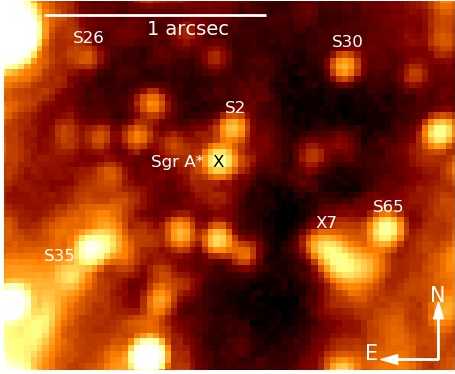}
\caption{NACO $L'$-band mosaic of the Galactic Centre in 2013.356. North is up and east is to the left. Identification of 3 stars (S26, S30 and S35) as photometric calibrators is depicted in particular. The black cross shows the infrared counterpart of Sgr~A*. In order to check any variability in the flux of S2, we used S65 as a reference star.}
\label{calibrators}
\end{figure}

The magnitude calibration is done by considering S26, S30 and S35  as calibrators using their known brightness from \cite{Gillessen2009}. These S-stars are chosen because of being isolated enough over the epochs of our study. In Fig.~\ref{calibrators}, we show the surroundings of S2 within about $\pm1\arcsec$ on 9 May 2013.

In general, the grainy background at the centre of our Galaxy is variable due to the proper motion of the stars \citep{2012AA...545A..70S}. Therefore, to correct for background brightness, we used five apertures with the same size as our photometric apertures close enough to the calibrators and S2 to get an appropriate representative of the existing background. Because of the high-velocity S-stars, the background is highly variable, therefore background apertures are not located at the same position for all the epochs considered in this study.

\subsection{Photometry results}
\label{Photometry results}

S2 is an early B-type dwarf of spectral type B0-B2.5 V with a stellar wind of estimated velocity $v_{\rm w} \sim 1000$ km s$^{-1}$ and a mass-loss rate of $\dot{m}_w \lesssim 3\times 10^{-7} M_{\odot} yr^{-1}$ \citep{2008ApJ...672L.119M}. We did the aperture photometry with local background subtraction for seven selected epochs from 2004 to 2018. S65 flux and magnitude was also studied in order to be a test star to reveal any meaningful changes in S2 flux or magnitude. 

\begin{figure}[ht!]
\centering
\includegraphics[width=90mm]{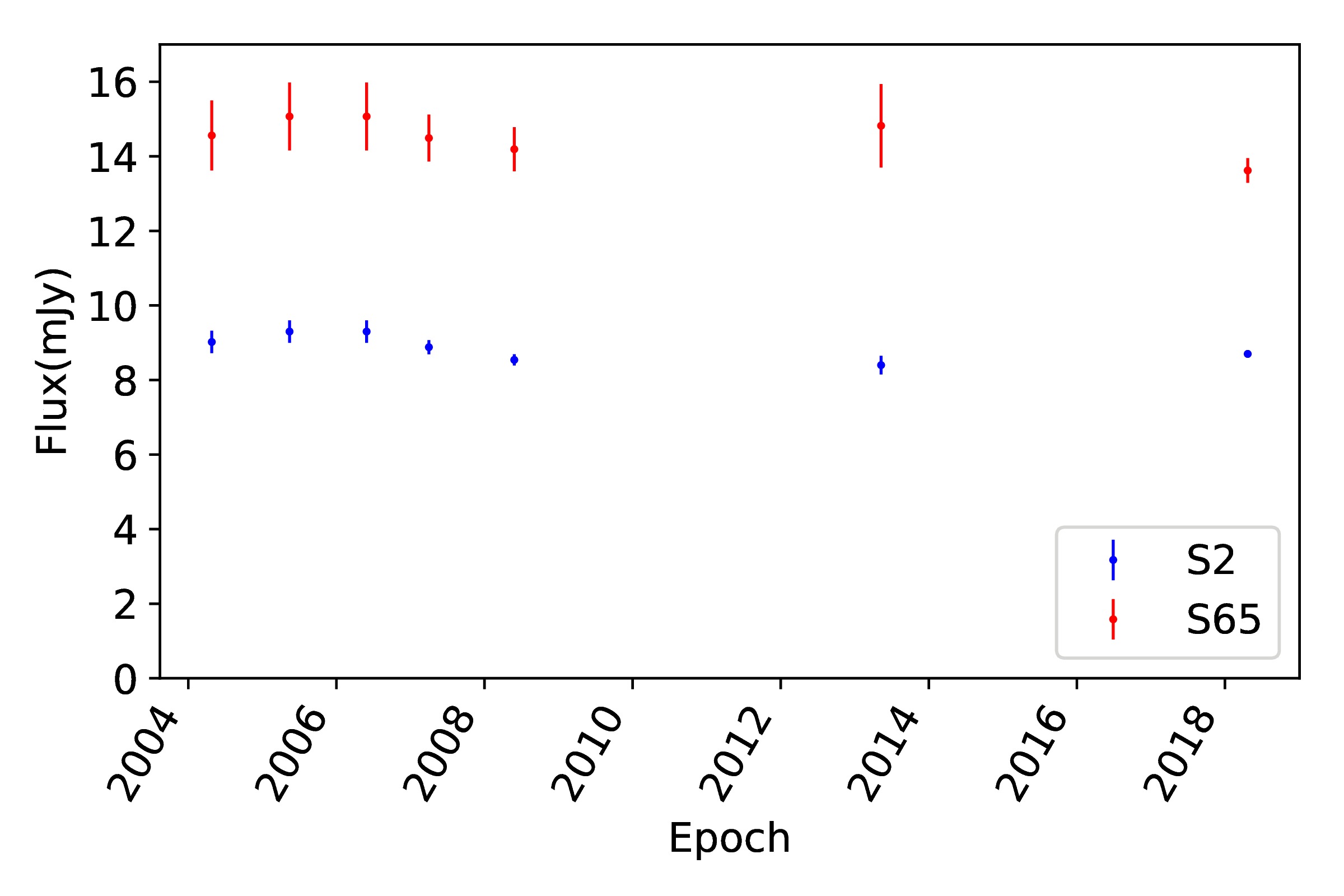}
\caption{Light curve of S2 and S65  in $L'$-band over seven observational epochs.}
\label{epochs-flux}
\end{figure}

The photometry results are presented in Table \ref{S2-epochs}. As it is shown in Fig. \ref{epochs-flux}, there is essentially no change detected in the measured flux density of S2 within the uncertainties over the seven epochs. The total coverage is 15 years, which is close to the S2 orbital period around Sgr~A*, including epochs close to the pericentre passages of S2.  The light curve has the mean value of the flux density of $\overline{F}=8.88 \pm 0.33\,{\rm mJy}$, with the difference of maximum and minimum values of $0.90\,{\rm mJy}$. In terms of the $L'$-band magnitude, the mean magnitude of S2 in the $L'$-band is $11.12 \pm 0.04$. The intrinsic variability of S2 is characterised by a normalized excess variance \citep{1997ApJ...476...70N},

\begin{equation}
    \sigma_{\rm rms}^2=\frac{1}{N\overline{f}^2}\sum_{i=1}^{N}[(f_i-\overline{f})^2-\sigma_i^2]\,,
    \label{eq_excess_var}
\end{equation}
where $f_i$ are individual flux measurements, $\overline{f}=1/N\sum_{i=1}^{N} f_i$ is the mean flux density, and $\sigma_i$ are individual measurement errors. We obtain the value of the fractional variability $F_{\rm var}=\sqrt{\sigma_{\rm rms}^2}=2.52\%$ in terms of the flux density and $M_{\rm var}=0.22\%$ in terms of the magnitude. For the control star S65, we obtain $F_{\rm var}=4.53\%$, which is almost a factor of two larger than S2. These values are fully consistent with no intrinsic changes in the $L'$-band flux density of S2 within measurement uncertainties. For completeness, the mean, the maximum difference, and the fractional variability $F_{\rm var}$ are listed in Table~\ref{S2-epochs} for both S2 and S65.

\begin{figure}[ht!]
\centering
\includegraphics[width=90mm]{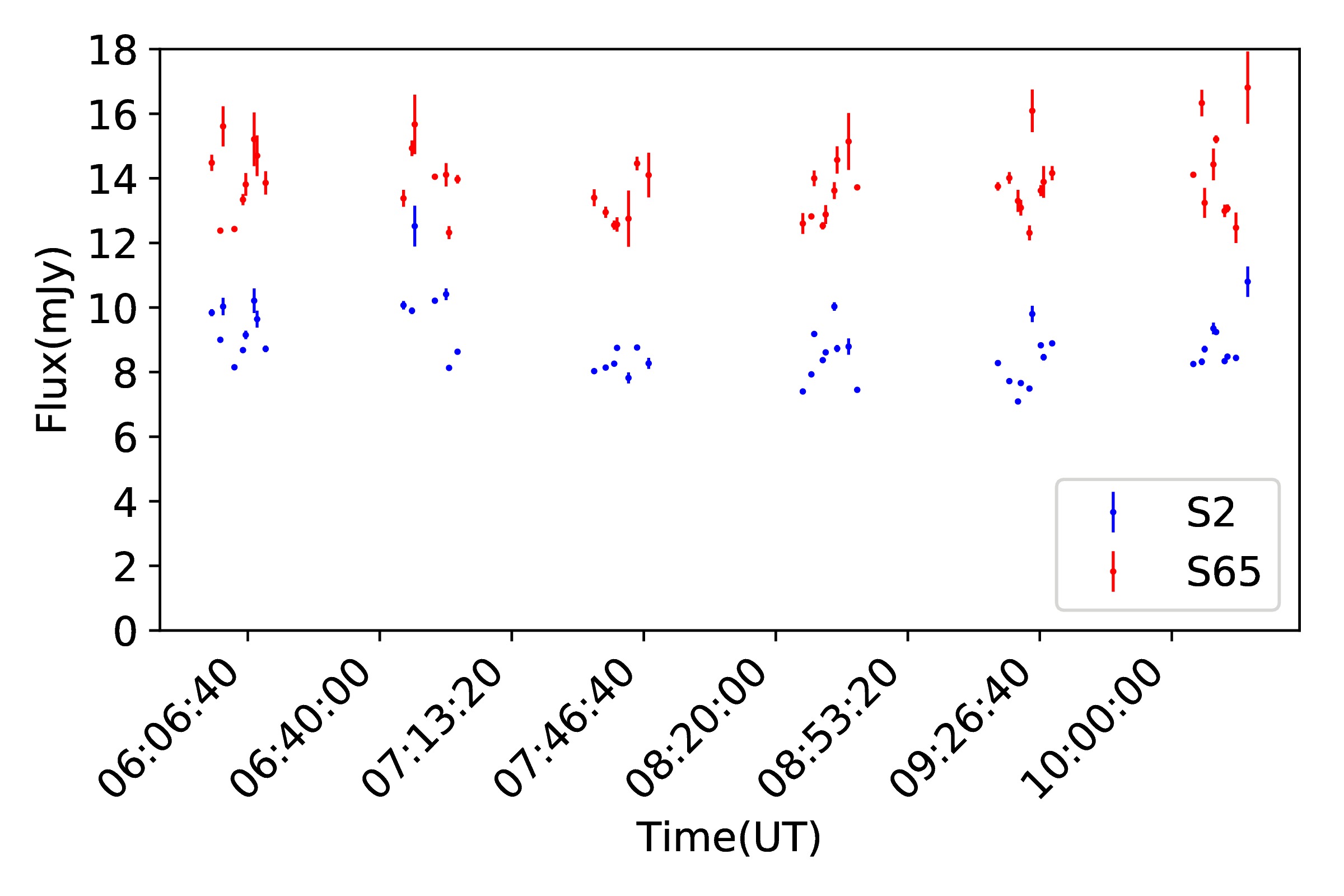}
\caption{Light curve at the position of S2 and the infrared counterpart of Sgr~A* and our reference star, S65, in 2018.307  in $L'$-band. The light curve shows the infrared counterpart of Sgr~A* is in its quiescence during our observation, therefore the light curve belongs to S2.}
\label{2018-flux}
\end{figure}

In order to study the S2 flux and magnitude closely during its periapse in May 2018 at around 1500 R$_s$ distance from Sgr~A*, we did the aperture photometry on all single exposures from the observation on April 22, 2018 on the position of S2 and the infrared counterpart of radio source Sgr~A*. The physical separation of S2 and Sgr~A* was only 1.38 mpc corresponding to 34.6 mas, while the angular resolution is 99.5 mas in $L'$-band. The photometry results are shown in Fig. \ref{2018-flux}. As it can be seen there is no sudden increase in the received flux from the position, therefore Sgr~A* is in its quiescent state in $L'$-band over the observation April 22, 2018. In addition to the non-detection of any flare from the infrared counterpart of Sgr~A*, no apparent increase in the S2 flux was observed within the measurement uncertainties. 

 For the 2002.663 epoch we measure 17.41 $\pm$ 1.18 mJy at the
- at that time - combined position of S2, G1 and $L'$-band
counterpart of Sgr~A*. Taking into account the S2's mean stellar flux density is 8.88 $\pm$ 0.33 mJy ( dreddened with $A_L = 1.09$) ,
the G1 $L'$-band flux density of $\sim 2.58 \pm 1.40 \,{\rm mJy}$ close to 2002 epoch \citep{Witzel_2017}, the flux of Sgr~A*
ends up to be around $6.03 \pm 0.52 $ mJy.
\cite{Schoedel-mean-flux-2011} obtains the mean flux of Sgr~A*
in the $L'$-band as $4.33 \pm 0.18 $  mJy, which leads us to $1.70$ mJy surplus in
the measured flux density.
However, due to the lack of a longer (at least days) light curve time coverage
during the S2 periapse passage of the variable source black hole counterpart
SgrA* in 2002, it is not possible to safely state that this flux density
excess is due to a Sgr~A* flare or the S2 stellar bow-shock effect. However, since during the subsequent pericentre passage in 2018 the flux density of S2 is within the uncertainty comparable to its mean value (see Table~\ref{S2-epochs}), the excess can be attributed to the Sgr~A* flare, unless the accretion flow density changes dramatically by several orders of magnitude from one periapse to another. The flux excess of $1.70\,{\rm mJy}$ that corresponds to the magnitude change of $\sim 0.19$ mag would require the pericentre number density of $\sim 4.2\times 10^{10}\,{\rm cm^{-3}}$ according to the dust-extinction model we present in Subsection~\ref{max_thermal_bowshock_contribution}. Considering the synchrotron bow-shock emission (see Subsection~\ref{max_nonthermal_bowshock_contribution}), the required density at the pericentre is smaller, $n_{\rm a}\sim 2.1 \times 10^5\,{\rm cm^{-3}}$, but still larger by two orders of magnitude than expected for the RIAF density profile with $\gamma\approx 1$, $n_{\rm a}\sim 7.2\times 10^3\,{\rm cm^{-3}}$. Such a large density present only for epoch 2002 is unlikely. The excess flux of $1.70\,{\rm mJy}$ can thus be attributed to the $L'$-band flare of Sgr~A*. In the $L'$-band, Sgr~A* is permanently variable \citep{Schoedel-mean-flux-2011} with the flux densities in the range between $1\,{\rm mJy}$ and $10\,{\rm mJy}$ \citep{2005ApJ...635.1087G,2008A&A...492..337E,2009ApJ...698..676D}.

\begin{table*}[tbh]
 \centering
 \begin{tabular}{ccccccc}
   \hline
   \hline
UT Date & Decimal Date & S2 flux(mJy)  & S65 flux (mJy)&  S2 mag & S65 mag & Observation ID\\
   \hline
2004 April 25 & 2004.317	&	9.02    $\pm$	0.30		&	14.56	$\pm$	0.94	&	11.10	$\pm$	0.04		&	10.59	$\pm$	0.07 & 60.A-9026(A)\\
2005 May 14 & 2005.367	&	9.30	$\pm$	0.30		&	15.07	$\pm$	0.91	&	11.07	$\pm$	0.04		&	10.55	$\pm$	0.07 & 073.B-0085(D)\\
2006 May 29 & 2006.408	&	9.30	$\pm$	0.30		&	15.07	$\pm$	0.91	&	11.07	$\pm$	0.04		&	10.55	$\pm$	0.07 & 077.B-0552(A)\\
2007 April 01 & 2007.249	&	8.88    $\pm$	0.19		&	14.49	$\pm$	0.63	&	11.12	$\pm$	0.02		&	10.59	$\pm$	0.05 & 179.B-0261(A)\\
2008 May 26 & 2008.402	&	8.54    $\pm$	0.15		&	14.19	$\pm$	0.59	&	11.16	$\pm$	0.02		&	10.61	$\pm$	0.05 & 081.B-0648(A)\\
2013 May 09 & 2013.356	&	8.40	$\pm$	0.25		&	14.82	$\pm$	1.12	&	11.18	$\pm$	0.03		&	10.57	$\pm$	0.08 & 091.C-0159(A)\\
2018 April 22 & 2018.307	&	8.70	$\pm$	0.10		&	13.62	$\pm$	0.33	&	11.13	$\pm$	0.01    	&	10.64	$\pm$	0.03 & 0101.B-0052(B)\\
\hline
mean &    &$8.88 \pm 0.33$ & $14.55 \pm 0.48$  & $11.12 \pm 0.04$ & $10.59 \pm 0.03$  \\
difference (max - min) &   &$0.90$ & $1.45$  &  $0.11$ & $0.09$   \\
$F_{\rm var}=\sqrt{\sigma_{\rm rms}^2}\,(\%)$ &    &$2.52$ & $4.53$  & $0.22$ & $0.51$ \\   
   \hline
 \end{tabular}
 \caption{The  $L'$-band flux and magnitude of S2 and S65 for seven epochs over a 15-year-interval from April 2004 to April 2018. The detector integration time (DIT) for all cases is 0.200 second.  Light curve characteristics, namely the mean, the difference between the maximum and the minimum, and the fractional variability are also included. With the fractional variability of $2.52\%$, no significant difference is detected for S2 flux within the uncertainties. The reported fluxes are dereddened.}
 \label{S2-epochs}
\end{table*}

\section{Constraining the slope and density of the ambient accretion flow using S2's $L'$-band emission}
\label{observability}

 In this section, we develop a simple method of inferring the density slope and the maximum density at the pericentre of a wind-blowing star using the observed light curve and an analytical dust extinction model. This method is suitable for stars whose bow shock is unresolved. For a fully resolved bow shock at at least two positions of a star along an elliptical orbit, one can infer the density slope from the bow-shock size ratio and an orbital eccentricity, see Appendix~\ref{derivation} for more details. Furthermore, we also consider the non-thermal bow-shock emission as an independent way of constraining the ambient density.
The S2 star is currently intensively monitored by the Very-Large-Telescope Interferometer GRAVITY, as well as the NIR-imager NACO and  SINFONI@VLT, which is capable of performing integral field spectroscopy.  Using the data over the past 25 years, post-Newtonian effects were measured, including the gravitational redshift \citep{2018AA...615L..15G} and the Schwarzschild precession \citep{2020AA...636L...5G,Parsa2017}, which are consistent with general relativity so far. Its highly eccentric orbit $(e=0.885)$ implies that the ambient gas density is expected to change along its orbit -- it is expected to be the smallest close to its apocentre and the largest close to the pericentre due to the radial power-law profile of the hot X-ray atmosphere \citep{2013Sci...341..981W}. 
The dust component is expected to coexist to a certain extent in this hot environment, which is also manifested by the presence of $L'$-band dusty sources in the central $\sim 0.04\,{\rm pc}$, namely DSO/G2, G1, and several other dusty and bow-shock sources in the central arcsecond \citep{2012Natur.481...51G,2015ApJ...800..125V,2017ApJ...847...80W}. In particular, for the orbital solutions and radiative properties of the dust-enshrouded objects, see \citet{peissker2020} and \citet{2020Natur.577..337C}.  
\begin{figure}[!ht]
    \centering
    \includegraphics[width=0.5\textwidth]{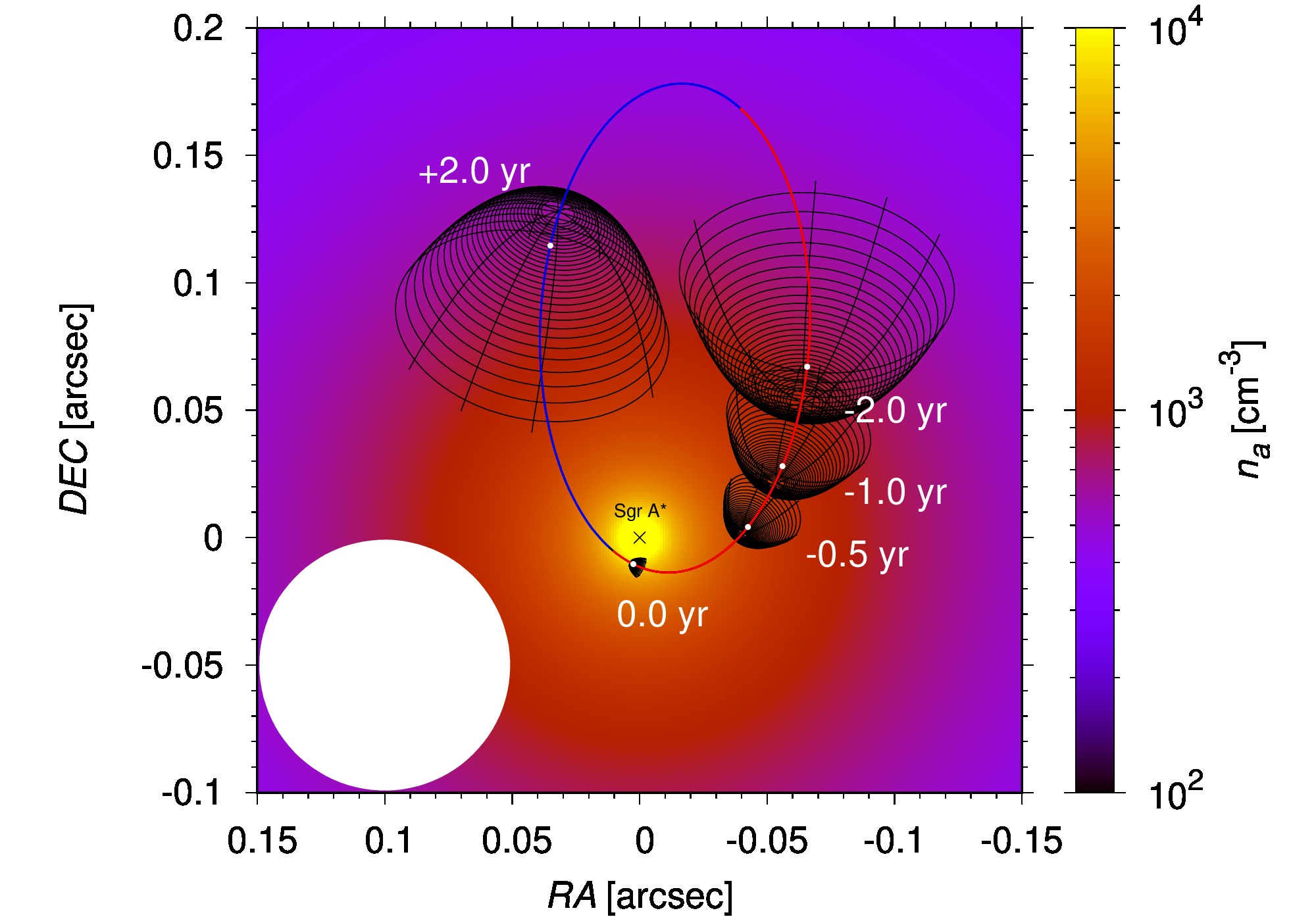}
    \caption{An illustrative figure of the S2-star motion through the hot flow around Sgr~A* of variable density. The density may change by as much as an order of magnitude from the apocentre to the pericentre, from several hundred particles per cubic centimetre to several thousand particles per cubic centimetre, respectively, depending on the power-law slope of the radial number density distribution.  For the exemplary calculation, we used $n_{\rm a}=n_{\rm B}(r/r_{\rm B})^{-1}$, where $n_{\rm B}=20.3\,{\rm cm^{-3}}$ is the particle density at the Bondi radius. The  white circle in the bottom left corner depicts the $L'$-band  point spread function (PSF) corresponding to 8-m class telescopes, from which it is apparent that the S2 bow-shock remains unresolved, especially close to the pericentre.}
    \label{fig_S2_motion}
\end{figure}
 An increase in the ambient gas-and-dust density along with the change in the orbital velocity are expected to lead to a larger local extinction around S2 star due to the formation of a bow shock, and hence to a change in the NIR magnitudes and corresponding colour indices. We considered here in particular the NIR $L'$-band emission of S2, which can trace such a change as S2 moves through the accretion flow of different gas and dust density, see Fig.~\ref{fig_S2_motion} for an illustration. A potential formation of a stellar bow-shock associated with S2 can enhance the colour change despite no sufficient capability of current NIR instruments to resolve the bow-shock structure, see Fig.~\ref{fig_S2_motion}.
 Both the detection as well as the non-detection of the $L'$-band magnitude can be used to constrain the density as well as the slope of the gas-and-dust density distribution. The density changes along the S2 orbit depends on the density slope $\gamma$. Assuming the radial gas distribution, $n_{\rm a}\approx n_0(r/r_0)^{-\gamma}$, where $\gamma>0$, the density ratio between the pericentre and the apocentre can be estimated as,
 \begin{equation}
    \frac{n_{\rm a,P}}{n_{\rm a, A}}=\left(\frac{r_{\rm A}}{r_{\rm P}}\right)^{\gamma}=\left(\frac{1+e}{1-e}\right)^{\gamma}\,.
    \label{}
 \end{equation}
 Since the orbital eccentricity is well constrained for S2, $e=0.885$, the ratio can be estimated to be $n_{\rm a,P}/n_{\rm a,A}\sim 4.05$, $16.39$, and $66.36$ for the plausible density slopes of $0.5$, $1.0$, and $1.5$, where the first two values represent the radiatively inefficient accretion flow \citep[see e.g.][]{2006ApJ...640..319X,2013Sci...341..981W} and the last value corresponds to the spherical steady Bondi flow \citep{2015AA...581A..64R}. Since the variation in density is large (by a factor of $16.4$) for a rather small variation in the density slope (by a factor of three), the $L'$-band observations of S2 can be used to constrain the density distribution of the hot accretion flow close to Sgr~A* at the length-scales of the order of 1500 Schwarzschild radii, where there are essentially no constraints for the ambient density \citep[see, however,][]{2019ApJ...871..126G}.\\  
 
In the further model, we approximate the orbit of S2 with an ellipse with the semimajor axis of $0.12538'' \approx 5\,{\rm mpc}$ and the eccentricity of $0.88473$ according to recent GRAVITY measurements \citep{2018AA...615L..15G}. These orbital elements imply that S2 covers a large distance range, with the pericentre distance of $r_{\rm p}=119.2\,{\rm AU}=1512\,r_{\rm s}$ and the apocentre distance of $r_{\rm a}=(1+e)/(1-e)r_{\rm p}=16.35r_{\rm p}=24724\,r_{\rm s}$. 
 \begin{figure}[ht]
     \centering
     \includegraphics[width=0.49\textwidth]{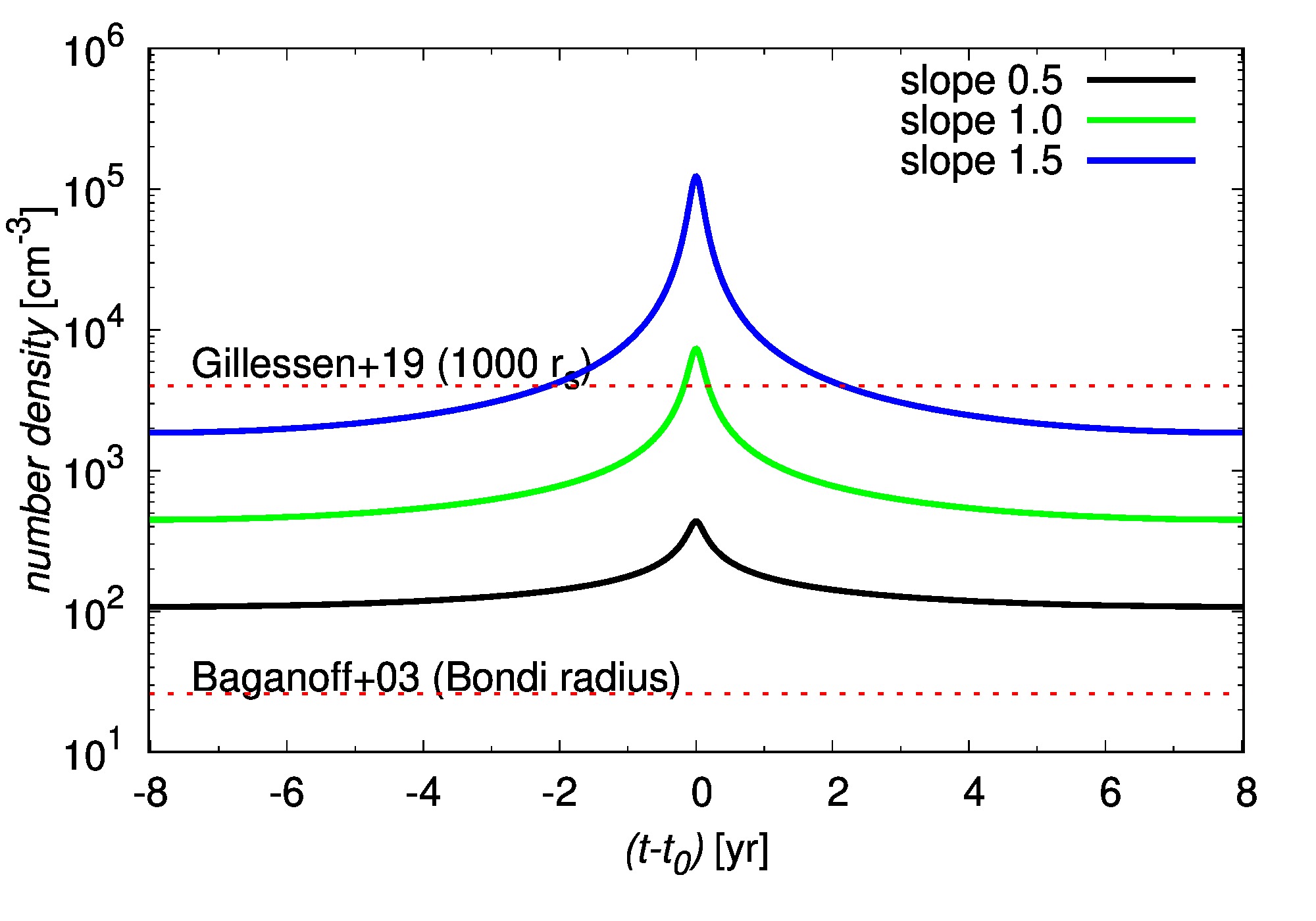}
     \caption{Temporal evolution of the number density of the ambient medium for the three different density slopes of the radial density profile, $n_{\rm a}=n_0(r/r_0)^{-\gamma}$, where $\gamma$ is equal to $0.5$, $1.0$, and $1.5$, see the plot legend. The dashed red lines stand for the number densities that were observationally inferred by \citet{2003ApJ...591..891B} (26 ${\rm cm}^{-3}$ at the Bondi radius) and \citet{2019ApJ...871..126G} (4000 ${\rm cm}^{-3}$ at $\sim 1000\, r_{\rm s}$).}
     \label{fig_density_ratio}
 \end{figure}
 The power-law density distribution is adopted according to the Chandra X-ray studies of the extended emission of Sgr~A* in the X-ray domain \citep{2003ApJ...591..891B,2013Sci...341..981W}. The density distribution can be scaled with respect to the Bondi capture radius, $n_{\rm a}\approx n_{a,B}(r/r_{\rm B})^{-\gamma}$, where $n_{\rm a, B} \approx 26 \eta_{\rm f}^{-1/2}\,{\rm cm^{-3}}$ and $r_{\rm B}\approx 4''(T_{\rm a}/10^{7}\,{\rm K})^{-1}=0.16\,{\rm pc}$  \citep{2003ApJ...591..891B,2013Sci...341..981W}. In Fig.~\ref{fig_density_ratio}, we show the number gas density per cubic centimeter that is expected to be intercepted by S2 star during its entire orbit (zero time stands for the orbit periapse). Different lines stand for different density slopes $(0.5;1.0;1.5)$ according to the key. The bottom dashed red line depicts the inferred number density close to the Bondi radius according to X-ray observations of \citet{2003ApJ...591..891B}, while the upper dashed red line stands for the number density of $n_{\rm a}^{DSO/G2}\sim 4 \times 10^{3}\,{\rm cm^{-3}}$ inferred for the periapse of the DSO/G2 object \citep{2019ApJ...871..126G}. 
 
\begin{figure}[!ht]
     \centering
     \includegraphics[width=0.5\textwidth]{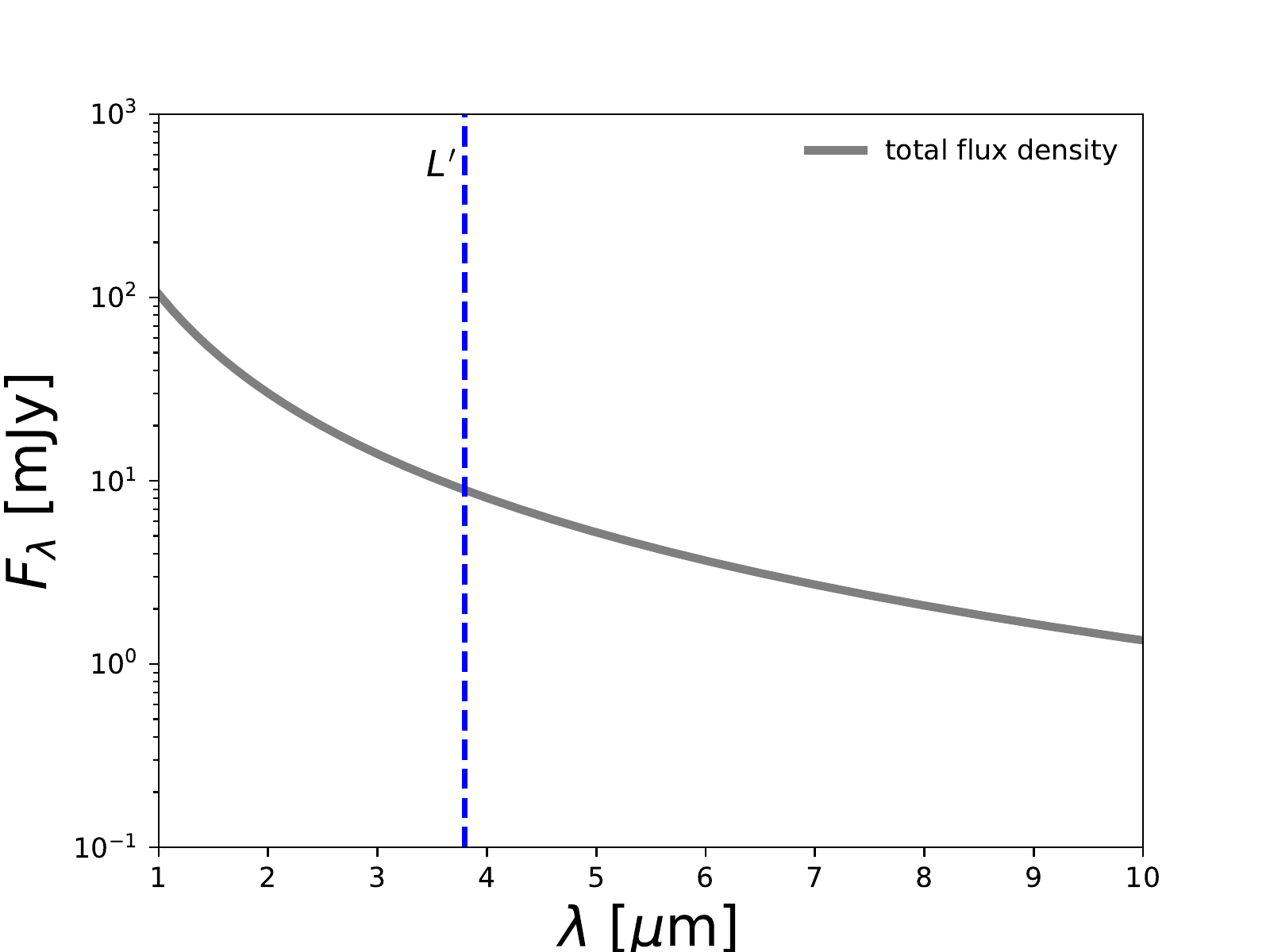}
     \caption{Total SED of the S2 star (black solid line) around $L'$ band.}
     \label{fig_SED_S2}
 \end{figure}

 We see that due to an eccentric orbit, S2 can in principle serve as a good probe of the ambient gaseous-dusty medium.
 For the S2 stellar blackbody emission, we adopt the stellar parameters that can reproduce the $L'$ flux density of $\sim 9\,{\rm mJy}$. For the model with $\log{(L_{\rm S2}/L_{\odot})}=4.54$, $T_{\rm S2}=26800\,{\rm K}$, $R_{\rm S2}=6\,R_{\odot}$, we obtain the $L'$ flux density of $F_{L}=8.9\,{\rm mJy}$, see also Fig. \ref{fig_SED_S2} for the spectral energy distribution (SED) of S2 around $L'$-band.
\subsection{Extinction estimate inside the Bondi radius}
\label{extinction} 
 We assume that the ambient gas inside the Bondi radius has a power-law profile, $n_{\rm a}\approx n_{\rm a,B}(r/r_{\rm B})^{-\gamma}$, which we scaled to the number density at the Bondi radius $r_{\rm B}$. To estimate the local extinction due to the gas and dust inside the Bondi radius, we calculate the local column density of hydrogen, 
 \begin{equation}
 N_{\rm h}=\int_{r_{\rm S2}}^{r_{\rm B}} n_{\rm a}(r)\mathrm{d}r=\frac{n_{\rm a,B}r_{\rm B}^{\gamma}}{1-\gamma}(r_{\rm B}^{1-\gamma}-r_{\rm S2}^{1-\gamma})\,,
 \end{equation}
 where we integrated from a certain inner radius associated with S2 star, in this case its pericentre distance of $r_{P}=a_{\rm S2}(1-e_{\rm S2})\approx 0.577\,{\rm mpc}$, up to the Bondi radius. The column density estimate for the number density of $n_{\rm B}\approx 26\,{\rm cm^{-3}}$ \citep{2003ApJ...591..891B} at the Bondi radius and the maximum density power-law slope of $\gamma\approx 1.5$ for the Bondi-type flow is $N_{\rm h}(\gamma=1.5)=4.02\times 10^{20}\,{\rm cm^{-2}}$, whereas for the flat slope of $\gamma=0.5$, we get $N_{\rm h}(\gamma=0.5)=2.41\times 10^{19}\,{\rm cm^{-2}}$. The visual extinction due to the ambient gas inside the Bondi radius for the steep density profile ($\gamma=1.5$) is given by $A_{\rm V}\approx 5.6\times 10^{-22}N_{\rm h}(\mathrm{cm}^{-2})=0.23\,{\rm mag}$ \citep[see e.g.][]{1973AA....26..257R,1975ApJ...198...95G,1995AA...293..889P}, which corresponds to the extinction $A_{\rm K}\sim 0.1\,A_{\rm V}=0.023\,{\rm mag}$ in the NIR $K$-band and $A_{\rm L}\approx 0.5\,A_{\rm K}=0.012\,{\rm mag}$ in the NIR $L'$-band \citep[see][for the extinction analysis towards the Galactic Centre]{2010AA...511A..18S}, in which our observations were carried out. For the flat density profile of $\gamma=0.5$, we get $A_{\rm L}\approx 0.0007$ mag. Hence, no increase in the $L'$-band magnitude within uncertainties is expected from the S2 plunging in deeper towards higher ambient densities. In summary, the upper limit on the ambient extinction inside the Bondi radius is $A_{\rm L}\lesssim 0.012\,{\rm mag}$. The only substantial increase at longer NIR wavelengths can be produced locally due to the bow-shock formation characterised by its density, radial scale, and thickness, which allows us to estimate the extinction.
 \subsection{Constraining the ambient density based on the thermal bow-shock emission}
 \label{max_thermal_bowshock_contribution}
  \begin{figure*}[tbh]
     \centering
     \includegraphics[width=0.48\textwidth]{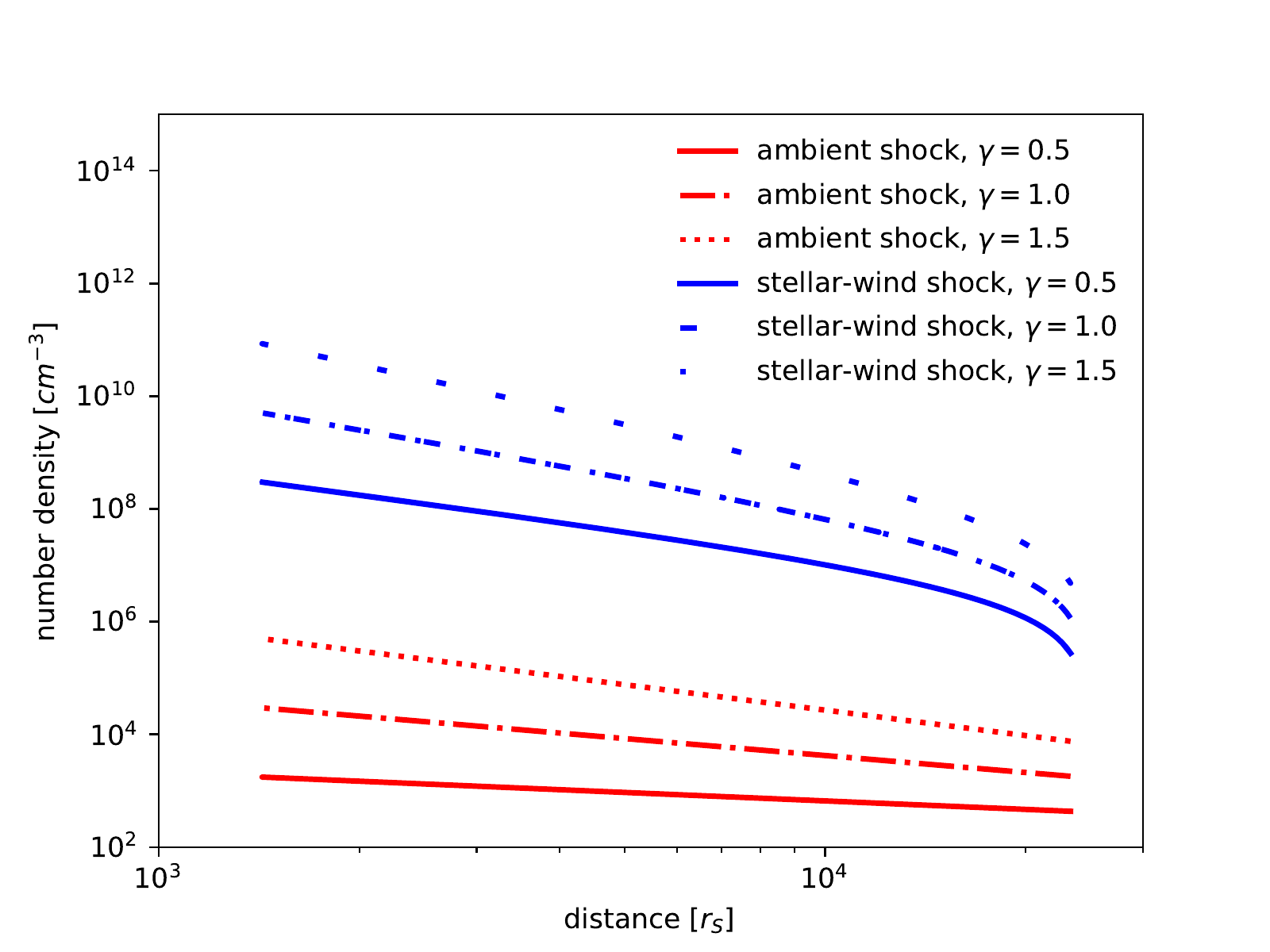}
     \includegraphics[width=0.48\textwidth]{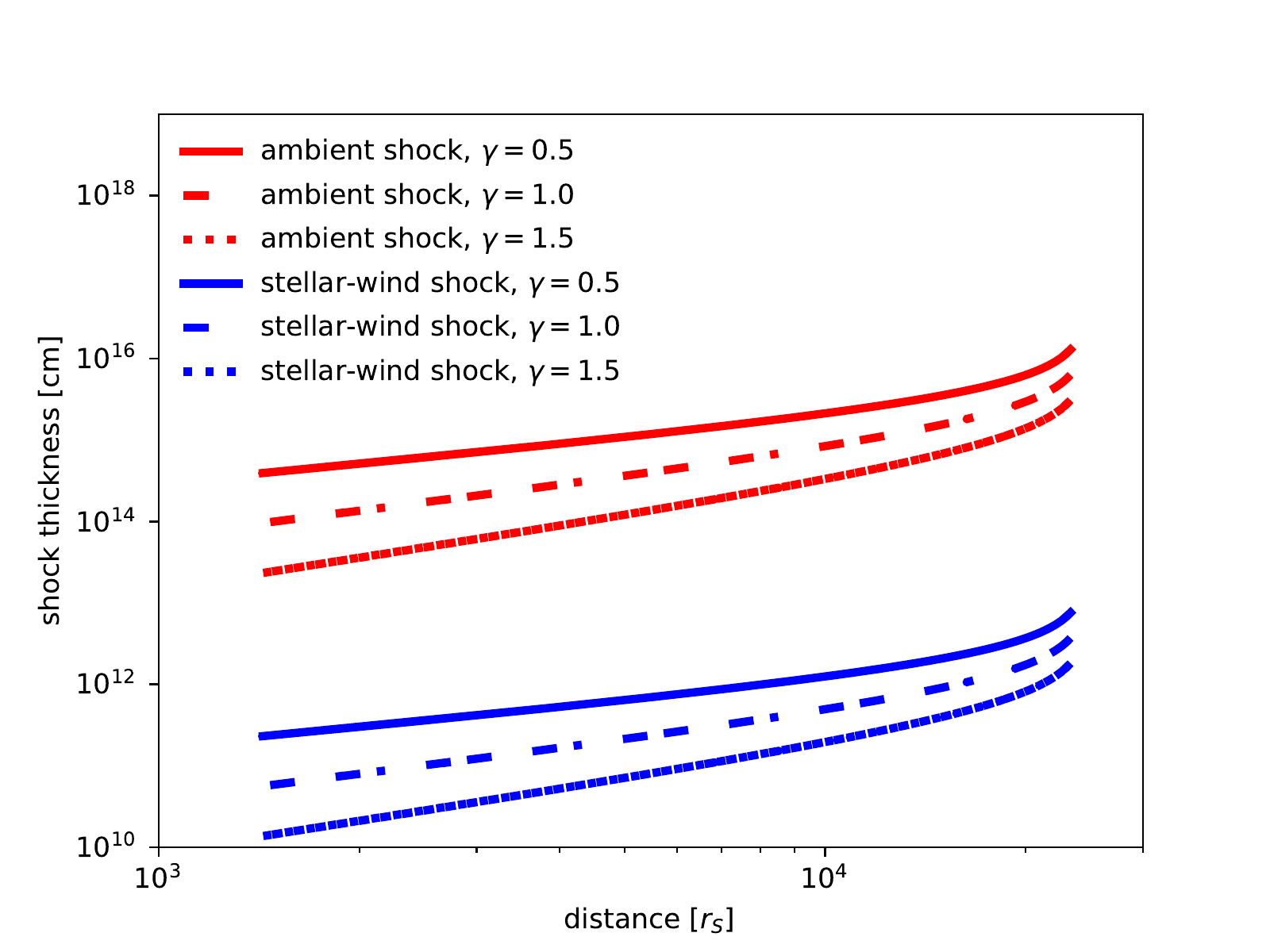}
     \caption{Number density and thickness of the two shock layers in the two-shock approximation model as a function of the distance from Sgr~A* (in Schwarzschild radii). \textbf{Left figure:} Number density (expressed in cm$^{-3}$)  of the shock driven into the ambient medium (red lines) and of the shocked stellar wind (blue lines) calculated for three specific values of the ambient density slope: $\gamma=0.5$, $1.0$, and $1.5$. \textbf{Right panel:} Thickness (expressed in cm) of the stellar-wind (blue lines) and ambient shock layers (red lines) calculated for the three different density slopes as in the left figure.}
     \label{fig_two_shocks}
 \end{figure*}
The S2 bow shock is modelled using a two shock scenario \citep{1975ApSS..35..299D}, where one shock is driven into the ambient medium (ambient shock) and the other is formed from the shocked stellar wind (stellar-wind shock). These two layers are separated by a contact discontinuity at the distance approximately given by the stagnation radius, see Eq.~\eqref{eq_stagnation_radius}. This treatment is based on the analytical theory of \citet{1975ApSS..35..299D}, who modelled an interaction of a fast stellar wind with an expanding gas globule. In the first approximation, two layers (warm and cold) mix to form one shocked layer. In the second approximation, two layers are separated by a contact discontinuity. We apply the second approximation in a similar way as \citet{2013ApJ...768..108S} did for the Galactic Centre NIR-excess source DSO/G2 interacting with the ambient diluted medium (see also their Fig. 3 for the illustration).
The extinction estimate is based on the number density of gas and dust inside the shock and its characteristic length-scale -- thickness. At the stagnation point, the ambient shock is characterised by the number density $n_{\rm ams}=4n_{\rm a}$ and the thickness of $t_{\rm ams}=0.65R_{0}$. The stellar wind shock has a characteristic number density of $n_{\rm sws}=n_{\rm w} v_{\rm w}^2/c_{\rm s}^2$, where $n_{\rm w}=\dot{m}_{\rm w}/4\pi R_0^2 v_{\rm w} \mu m_{\rm H}$ is the stellar wind number density with the mean molecular weight of $\mu=0.5$ (ionised gas) and the hydrogen mass of $m_{\rm H}$ and the sound speed of the stellar wind gas $c_{\rm s}=(k_{\rm B}T_{\rm sw}/\mu m_{\rm H})^{1/2}$ is evaluated for the temperature of $T_{\rm sw}=10^{4}\,{\rm K}$. The thickness of the stellar wind is given by $t_{\rm sws}=2.3 (c_{\rm s}/v_{\rm w})^2 R_0$. We show both the number density and the thickness of both shocks in Fig.~\ref{fig_two_shocks} as a function of the distance of S2 from Sgr~A* in Schwarzschild radii. We see that the ambient shock is thicker and less dense than the stellar wind shock for all ambient density profiles considered in the legend. 

In our model, we further consider that the dust is more likely to exist in the ambient medium through which S2 moves along its orbit, hence the ambient shock is expected to be more relevant in terms of the localized extinction and its contribution in the $L'$-band. The B-type stars typically do not contain dust in their stellar winds, hence the stellar wind shock is not likely to contribute to $L'$-band emission. However, for completeness, we consider both shock layers in assessing the constraints on the ambient number density. This is also due to the complex dynamics of dust grains and their potential of being dragged by a denser material in the stellar-wind shock, especially smaller grains may be affected \citep[see e.g.][]{2011ApJ...734L..26V}. During the motion of S2, the two shock layers are affected by hydrodynamic instabilities that lead to the mixing of both layers \citep{Schartmann+2018}, which is likely to lead to the exchange of material and the actual density of the gaseous-dusty mixed layer may be closer to the denser stellar-wind shock.

We calculate the limits on the ambient number density at the S2 periapse as well as the limits for the density slope $\gamma$. The upper limit is obtained if we associate the upper limit for the extinction with the standard deviation, $A_{\rm shock}\lesssim 0.04$ mag. Another estimate is based on the  fractional variability with $A_{\rm shock}\sim 0.02$ mag. The values of the density and the slope are inferred from the extinction-column density relation $A_{\rm V}\approx 5.6\times 10^{-22}N_{\rm h}(\mathrm{cm}^{-2})$, where $N_{\rm h}\sim n_{\rm shock}t_{\rm shock}$ with the shock number density $n_{\rm shock}$ and the shock thickness $t_{\rm shock}$ for each shock layer as explained above. The extinction in $L'$-band is approximately given by $A_{L}=0.5\times 0.1 \times A_{V}$ \citep{2010AA...511A..18S}. This model allows us to compute the whole light curve along the S2 trajectory. In Table~\ref{tab_density}, we list the density slope and the number density constraints for each extinction value as well as the shock layer taken into account. For the ambient shock, which is more likely to lead to an increased dust emission, the upper limits on the number density are in the range $n_{\rm a}^{\rm ams}\lesssim 7.15\times 10^8-1.87 \times 10^9\,{\rm cm^{-3}}$ with the density slope $\gamma \lesssim 3.03-3.20$. For the stellar-wind shock, the number densities are smaller by four orders of magnitude and the density slope is smaller by a factor of two, $n_{\rm a}^{\rm sws}\lesssim 7.53 \times 10^4-1.86 \times 10^5\,{\rm cm^{-3}}$ and $\gamma \lesssim 1.41-1.57$. 

\begin{figure*}
    \centering
    \begin{tabular}{cc}
    \includegraphics[width=0.5\textwidth]{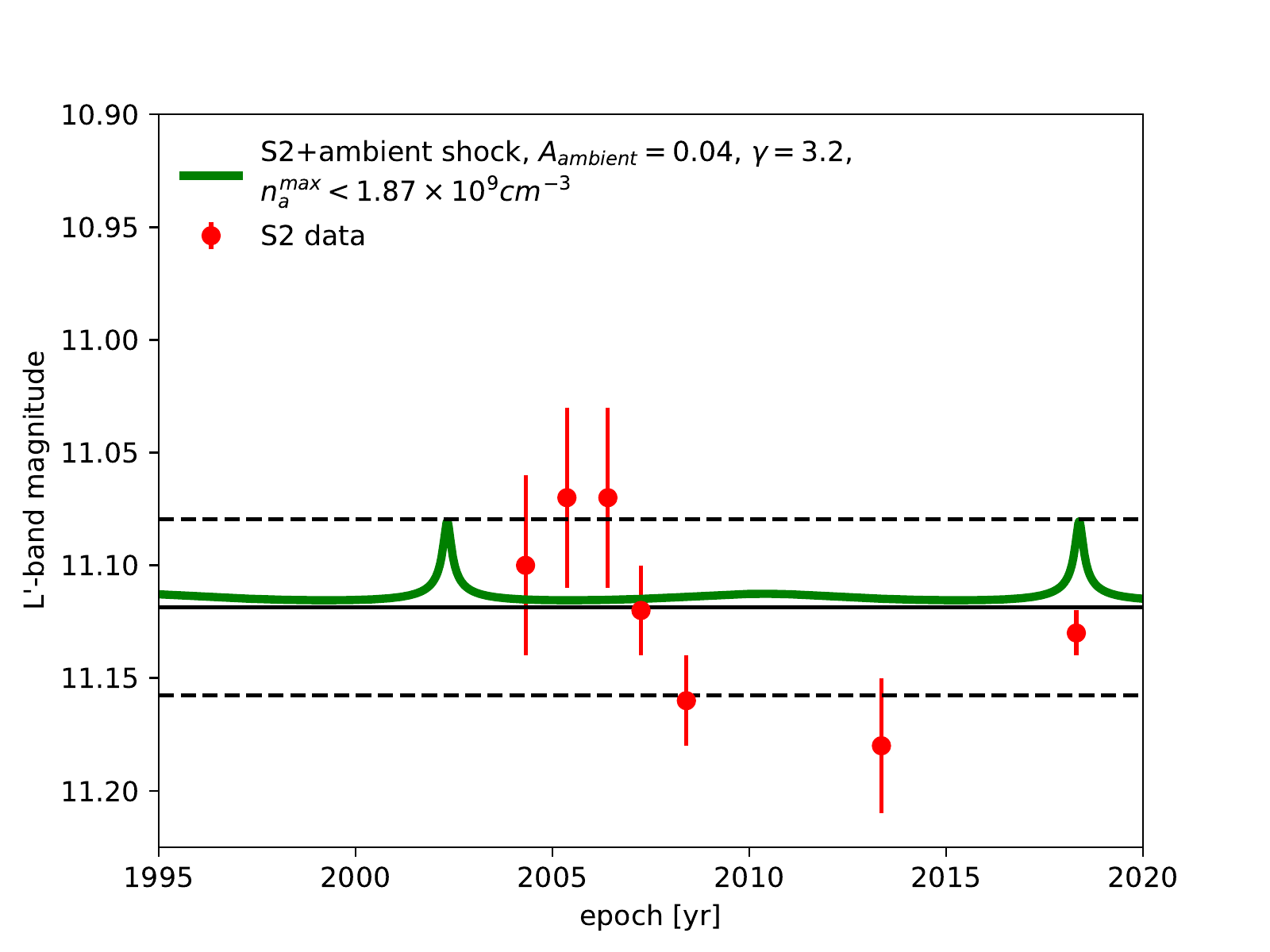}
    \includegraphics[width=0.5\textwidth]{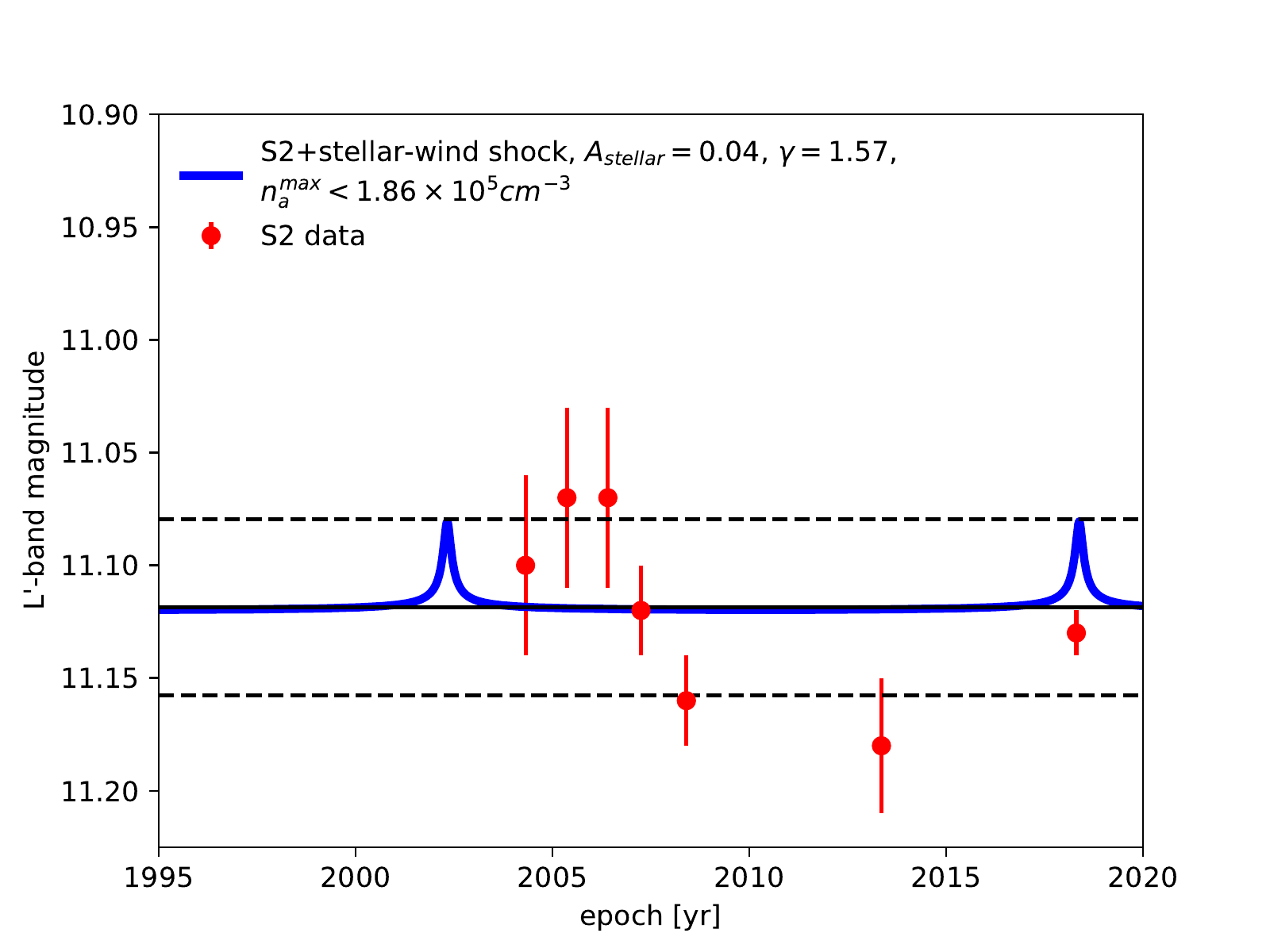}\\
    \includegraphics[width=0.5\textwidth]{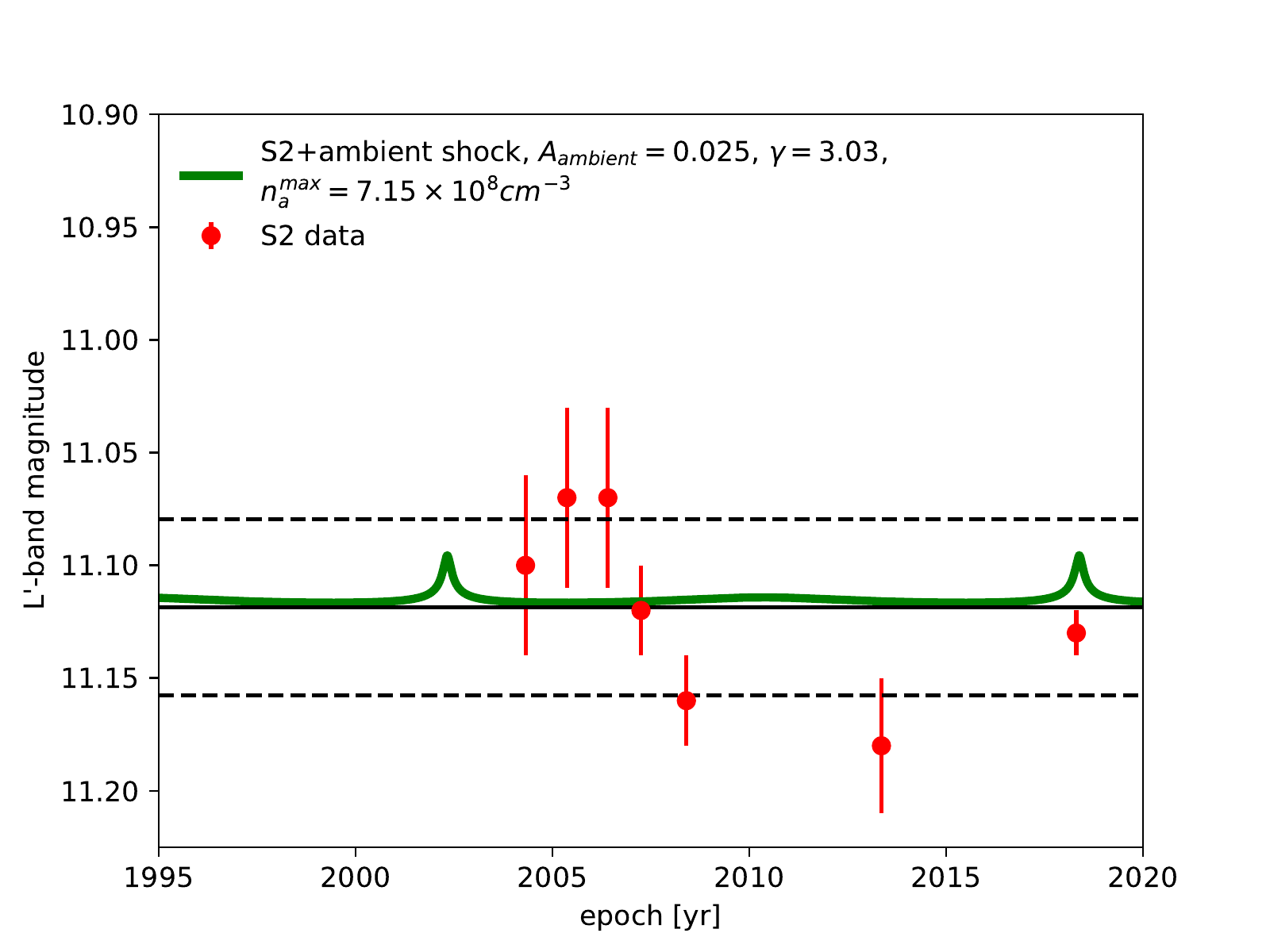}
    \includegraphics[width=0.5\textwidth]{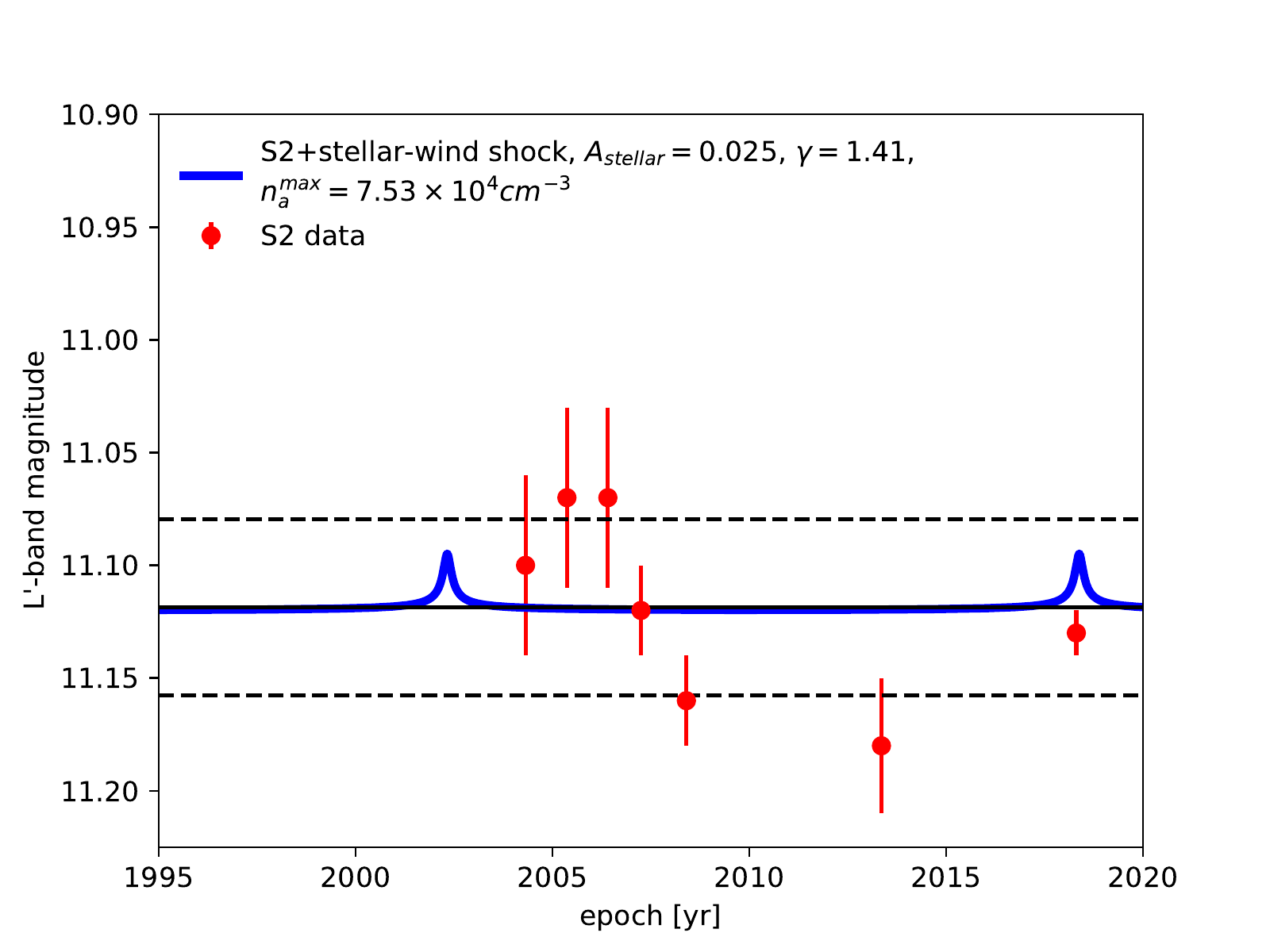}
    \end{tabular}
    \caption{Comparison of model light curves based on the mean S2 emission and the modelled thermal emission of its bow shock (solid green and blue lines) with the observed $L'$-band emission of S2 (red points with errorbars).  {\bf Upper panels}: The modelled thermal emission of the bow shock is associated with the extinction of $A=0.04$ mag, which was inferred from the standard deviation of observed light curve points. We model separately the thermal dust emission of the ambient bow shock in the left panel (solid green line) and the thermal emission of the stellar-wind shock is displayed in the right panel (solid blue line). The horizontal solid black line represents the light curve mean value, while the dashed black lines stand for the standard deviation.  {\bf Lower panels}: As in the upper panels, but for the extinction of $A=0.02$ mag based on the excess variance of the observed light curve points.}
    \label{fig_light_curve_mag}
\end{figure*}

The comparison of model light curves with the actual $L'$-band data is in Fig.~\ref{fig_light_curve_mag}, with the upper panels representing the larger extinction value of $A=0.04$ based on the standard deviation of the light curve points and the lower panels standing for the smaller extinction of $A=0.02$ based on the fractional variability. The peak of the observed light curve has an apparent offset from the peak of the modelled light curves (at the periapse), which is most likely a systematic effect, not an intrinsic brightening.   
\begin{table*}
    \centering
    \begin{tabular}{c|c|c}
        \hline
        \hline
        Thermal emission -- upper limit & power-law slope &  ambient number density (periapse) \\
         \hline
         ambient shock ($A=0.04$) & $3.20$ & $1.87\times 10^9\,{\rm cm^{-3}}$     \\
         stellar-wind shock ($A=0.04$) & $1.57$ &  $1.86 \times 10^5\,{\rm cm^{-3}}$    \\
         \hline 
        Thermal emission -- upper limit  &  power-law slope & ambient number density (periapse)\\
          \hline
         ambient shock ($A=0.02$) & $3.03$ &  $7.15 \times 10^8\,{\rm cm^{-3}}$\\
         stellar-wind shock ($A=0.02$) & $1.41$ & $7.53 \times 10^4\,{\rm cm^{-3}}$\\
         \hline 
         Non-thermal emission -- upper limit & power-law slope & ambient number density (periapse)\\
         \hline
         peak flux density of $0.91\,{\rm mJy}$ (maximum flux difference)  & $1.47$ & $1.01\times 10^5\,{\rm cm^{-3}}$\\
         peak flux density of $0.22\,{\rm mJy}$ (fractional variability) & $1.17$ & $1.88\times 10^4\,{\rm cm^{-3}}$\\
         \hline
         \hline
    \end{tabular}
    \caption{ Summary of constraints for the ambient density close to the Galactic Centre at the S2 periapse based on the two extinction limits, 1$\sigma$ and fractional variability, and analytical bow-shock models for both the ambient shock and the stellar-wind shock thermal emission. In addition, we include the ambient density upper limit for the non-thermal synchrotron emission based on the maximum flux difference of $0.90\,{\rm mJy}$ and the fractional variability of $0.2\,{\rm mJy}$.}
    \label{tab_density}
\end{table*}
 \subsection{Constraining the ambient density based on the non-thermal bow-shock emission}
 \label{max_nonthermal_bowshock_contribution}
 
 The thermal bow-shock emission in $L'$-band is clearly affected by the presence of dust along the S2 orbit. While on one hand the dust is clearly present in dusty objects whose orbits lie in the S cluster, the hot ambient medium likely destroys the dust particles continually. In case the presence of dust is diminished along the S2 orbit, then the thermal emission of the bow shock in $L'$- band is even smaller than we predicted in Subsection~\ref{max_thermal_bowshock_contribution}.
 
 Another way to constrain the ambient density at the S2 pericentre is to take into consideration the broadband non-thermal synchrotron emission of the bow shock, which does not depend on the presence of dust. When the bow shock develops ahead of the star, it provides a volume where electrons can be accelerated by the enhanced magnetic field. Subsequently, they cool off by emitting synchrotron emission. \citet{Ginsburg+2016} used the same mechanism to claim that the non-thermal emission of stellar bow shocks in the S cluster may be comparable to the radio and near-infrared emission of Sgr~A*, and hence could be detected.
 
 One can reverse the task and instead constrain the upper limit of the ambient density based on the statistical variations we detect in the $L'$-band light curve of S2. According to Table~\ref{S2-epochs}, the maximum flux difference in our light curve is $0.90\,{\rm mJy}$, while the rms fractional variability is at the level of $0.2\,{\rm mJy}$. Hence, the maximum contribution of the bow-shock synchrotron emission in $L'$-band can only be at this flux density level, which allows us to place an upper limit on the ambient density.
 \begin{figure*}
     \centering
     \includegraphics[width=\columnwidth]{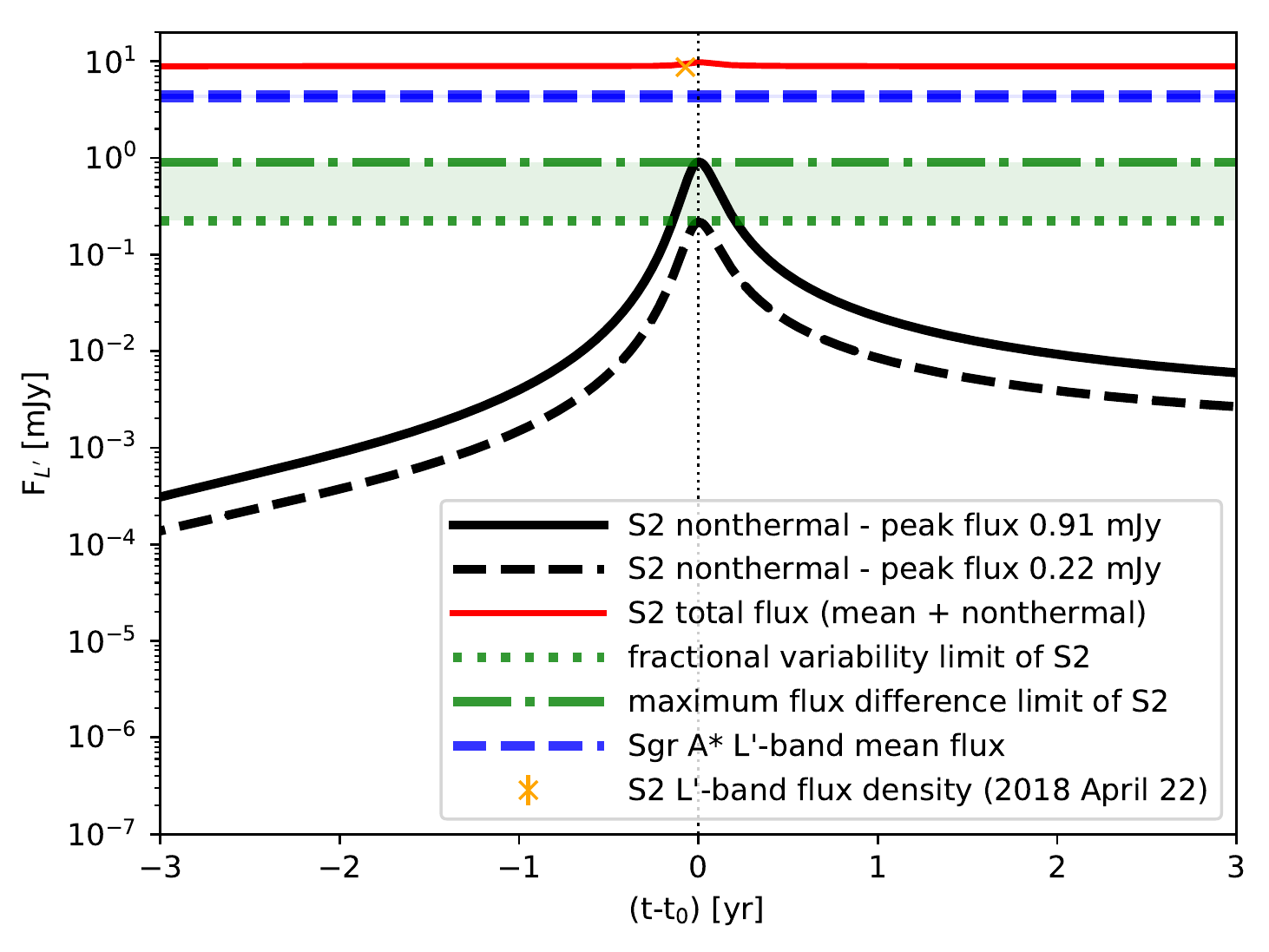}
     \includegraphics[width=\columnwidth]{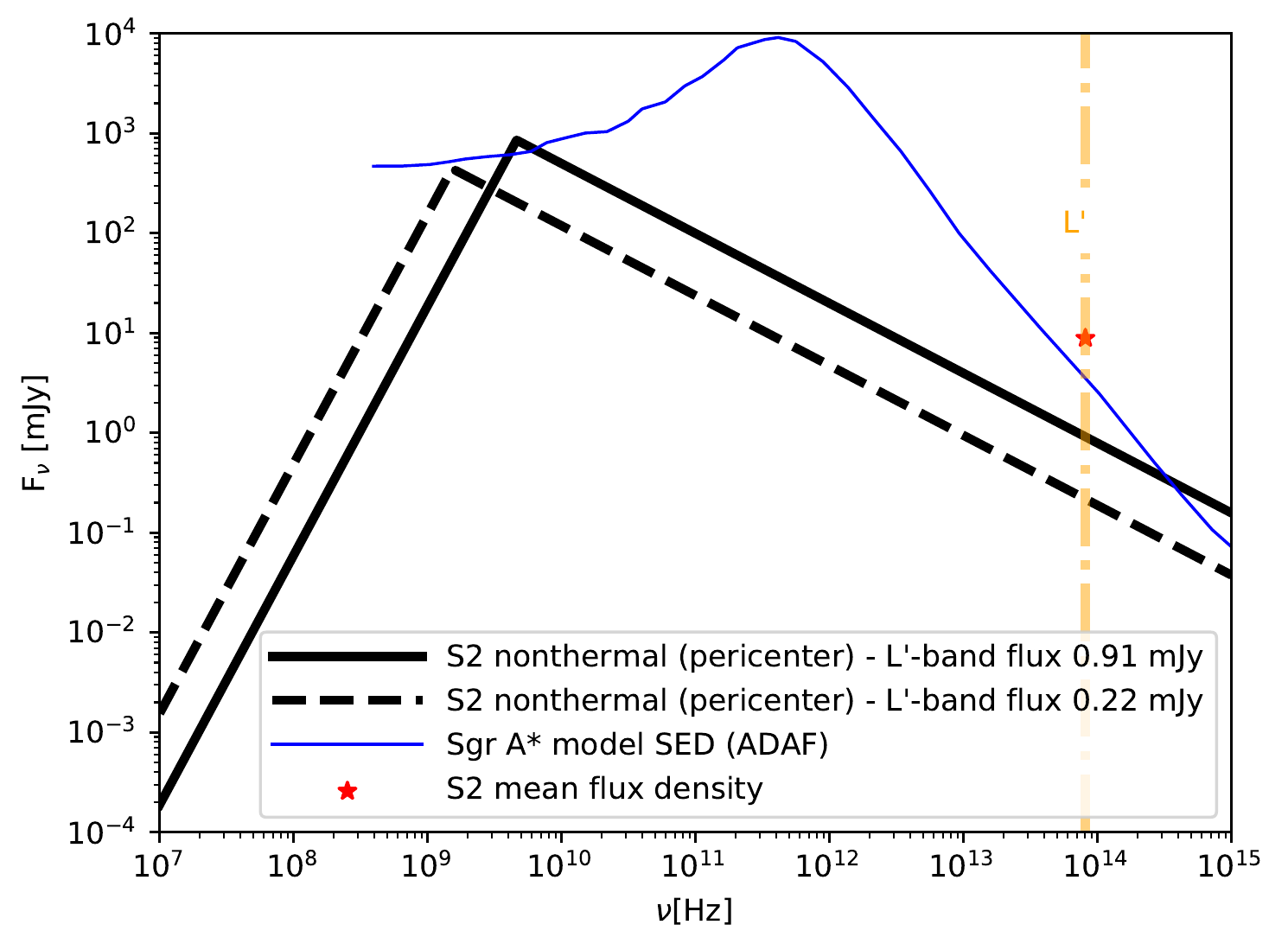}
     \caption{Nonthermal contribution of the S2 bow shock to its $L'$-band emission. In the left panel, we plot the $L'$-band light curve as calculated for the S2 bow-shock synchrotron emission for the peak flux of $0.91\,{\rm mJy}$ (solid black line) and for the peak flux of $0.22\,{\rm mJy}$ (dashed black line). For comparison, we also show the fractional variability of S2 (dotted green line) and its maximum flux difference (dot-dashed line). The measured mean flux of Sgr~A* in $L'$-band, $4.33\pm 0.18\,{\rm mJy}$, is depicted by a dashed blue line. The sum of the mean flux density of S2 with the maximum potential contribution of the nonthermal bow-shock emission is represented by a solid red line. The actual measured flux density of S2 from April 22 2018 is marked by an orange cross. In the right panel, we show the SED of S2 bow-shock synchrotron emission as calculated for the S2 pericentre for the $L'$-band flux of $0.91\,{\rm mJy}$ (solid line) and for the $L'$-band flux of $0.22\,{\rm mJy}$ (dashed line). The orange dot-dashed line marks the $L'$-band frequency range and the red star symbol represents the mean flux density of S2 in $L'$-band. The blue solid line represents the SED of Sgr~A* based on the advection dominated accretion flows \citep[ADAFs; ][]{2003ApJ...598..301Y,2014ARAA..52..529Y}.}
     \label{fig_S2_synchrotron}
 \end{figure*}
 
 For the calculation of the bow-shock synchrotron light curve and broad-band spectrum, we follow the model of \citet{2013MNRAS.432..478S}, originally developed for the estimate of the bow-shock emission of the DSO/G2 object, see also \citet{2016MNRAS.455.1257Z}. As in \citet{2013MNRAS.432..478S} and \citet{2016MNRAS.455.1257Z}, for calculating synchrotron light curves and spectra, we use the temperature profile of the ADAF, $T_{\rm a}=9.5 \times 10^{10}(r/r_{\rm s})^{-1}$ \citep{2003ApJ...598..301Y}, with the ratio between the magnetic and the gas pressure of $\chi=P_{\rm mag}/P_{\rm gas}=0.3$, and the Mach number of the star at the pericentre distance of S2, $\mathcal{M}=2.6$. Other relevant parameters were adopted from particle-in-cell simulations performed by \citet{2013MNRAS.432..478S}, specifically the slope of the high-energy power-law electron distribution, $p=2.4$, the minimum Lorentz factor, $\zeta=7.5$, and the fraction of accelerated electrons to high energies, $\eta=0.05$. Specifically for S2, we used its mass-loss rate $\dot{m}_{\rm w}\simeq 3\times 10^{-7}\,M_{\odot}\,{\rm yr^{-1}}$ and the terminal wind velocity $v_{\rm w}\sim 10^3\,{\rm km\,s^{-1}}$ inferred from spectroscopy \citep{2008ApJ...672L.119M,2017ApJ...847..120H}.

 In contrast to \citet{2013MNRAS.432..478S}, we perform only the plowing synchrotron model, in which the accelerated electrons are kept in the shocked region and radiate in the shocked magnetic field. The plowing model is more likely to contribute in $L'$-band than the local model, in which accelerated electrons leave the shock and radiate in the unshocked magnetic field. Due to the short cooling time for the near-infrared $L'$-band synchrotron emission, $t_{\rm cool}\approx 0.25(B/0.08\,{\rm G})^{-3/2}(\nu_{\rm c}/8 \times 10^{13}\,{\rm Hz})^{-1/2}\,{\rm yr}$, the local model would have a negligible contribution.
 
 In Fig.~\ref{fig_S2_synchrotron} (left panel), we plot the $L'$-band synchrotron light curve for the peak flux density of $0.91\,{\rm mJy}$, which corresponds to the maximum flux difference, and the light curve with the peak flux density of $0.22\,{\rm mJy}$, which corresponds to the fractional variability. The non-thermal bow-shock emission is also compared with the Sgr~A* mean flux density in $L'$-band \citep{Schoedel-mean-flux-2011}, the S2 total flux, and the actual S2 measured flux density from April 22, 2018, which are all larger by at least a factor of a few in comparison with the bow-shock non-thermal emission. In the right panel of Fig.~\ref{fig_S2_synchrotron}, we plot the SED of the S2 bow-shock synchrotron emission calculated for the pericentre for both $L'$-band flux density values (maximum difference and fractional variability). We also compare the synchrotron bow-shock SED with the SED of Sgr~A* calculated using the advection dominated accretion flow model \citep[ADAF; ][]{2003ApJ...598..301Y,2014ARAA..52..529Y}. Taking into account the maximum flux difference as an upper limit for the $L'$-band synchrotron contribution, we get the pericentre number density of $n_{\rm a}=1.01\times 10^5\,{\rm cm^{-3}}$ (the slope of $1.47$). The fractional variability flux limit yields the number density of $1.88\times 10^4\,{\rm cm^{-3}}$ (the density slope of $1.17$). Both limiting cases are summarised in Table~\ref{tab_density}. For both density limits, the bow-shock synchrotron flux density would be comparable to that of Sgr~A* at a few GHz frequencies, where the peak of its SED occurs.
To separate Sg~A* and the predicted S2 bow-shock emission is
challenging as it requires highly sensitive measurements
at cm- to dm-wavelengths with an angular resolution of a
tenth of an arcsecond or better.
However, around the time  of the periapse passage
longterm (weeks to a few months) light curves should reveal an overall
increase of the spatially unresolved emission of
both sources. Such a data set is not available at present. VLBI measurements at GHz frequencies (or cm wavelengths) could in principle be possible due to the high brightness temperature of the S2 bow-shock emission at its pericentre. We outline here a few basic estimates.  For the assumed maximum non-thermal excess in $L'$-band of $\sim 1$ mJy, the calculated peak of the bow-shock synchrotron spectrum is at $4.61$ GHz. The peak flux density is expected to be $0.86$ Jy at the S2 pericentre, which is comparable to or even larger than the flux density of Sgr~A* at cm wavelengths. The bow-shock characteristic size given by the stagnation distance would be $R_{0}\sim 5.6 \times 10^{13}\,{\rm cm}$ at the pericentre, which gives the angular scale of $\sim 0.5\,{\rm mas}$ that is below the resolution capabilities of the cm VLBI, which is at best $\sim 1\,{\rm mas}$. The angular distance between S2 radio source and Sgr~A* at the pericentre is $\sim 14.4$ mas, which is smaller than the intrinsic size of Sgr~A* at 6 cm ($\sim 20$ mas) due to $\lambda^2$ scatter-broadening of the size of Sgr~A* \citep{2006ApJ...648L.127B,2008Natur.455...78D}. The brightness temperature\footnote{We estimate the brightness temperature using the standard relation derived from the Rayleigh-Jeans approximation in the radio domain, $T_{\rm B}=\frac{c^2}{2k_{\rm B}}\frac{F_{\nu}}{\Omega \nu^2}$, where the solid angle $\Omega\approx \pi R_0^2/d_{\rm GC}^2$ with $R_0$ being the stagnation radius of the bow shock.} of the S2 bow shock at its pericentre would be large, $T_{\rm B}\sim 8.5\times 10^{10}\,{\rm K}$, comparable to that of Sgr~A*, $\gtrsim 2\times 10^{10}\,{\rm K}$ \citep{1998A&A...335L.106K,2008Natur.455...78D}. However, before and after the S2 pericentre, its brightness temperature is expected to drop rather fast, remaining above $10^6\,{\rm K}$ from the epoch $2017.08$ up to $2020.54$, with the peak at the pericentre in $2018.38$. Taking into account the smaller contribution of S2 bow-shock non-thermal emission at $L'$-band at the level of $\sim 0.2\,{\rm mJy}$, which results from a smaller ambient density, mainly the characteristic size is affected. It gets larger to $R_{0}=1.29\times 10^{14}\,{\rm cm}$, which corresponds to the angular scale of $1.1\,{\rm mas}$ that is at the limit of the resolving power of cm VLBI. However, since for a lower density the resulting spectrum of the S2 bow shock is shifted towards the lower frequency at $1.5$ GHz, see Fig.~\ref{fig_S2_synchrotron}, the bow-shock size is actually smaller than the typical VLBI beam FWHM of $\sim 3''$. With the peak flux of $435$ mJy, the brightness temperature of $T_{\rm B}\sim 7\times 10^{10}\,{\rm K}$ is again comparable to that of Sgr~A* and stays above $10^6\,{\rm K}$ in the time window between $2016.98$ and $2020.90$. In general, the cm VLBI observations could reveal a deviation from the Gaussian core of Sgr~A* in the epochs of $\sim 1.5$ years before and after the pericentre when the brightness temperature is still above $10^6\,{\rm K}$. At the pericentre, the S2 bow-shock emission is not possible to resolve out due to the scatter-broadening of Sgr~A* as well as of the bow-shock source. Only the flux excess could be detected, however, in that case a corresponding light curve at cm wavelengths would need to be analysed to distinguish the long-term bow-shock excess from the short-term stochastic radio flares.

 Based on the upper limit inferred from the maximum synchrotron constribution to the $L'$-band emission of S2, accretion flows exceeding the particle density of $\sim 10^5\,{\rm cm^{-3}}$ at $\sim 1500\,r_{\rm s}$ are unlikely. On the other hand, the presence of the Bondi-type flow with $\gamma\sim 3/2$ cannot be excluded based on our $L'$-band light-curve analysis.

\section{Discussion}
\label{Discussion}
Our observational analysis of the $L'$-band emission of the S2 star yielded a light curve that is essentially flat  within uncertainties and intrinsic flux variations due to the interaction of the S2 star with the ambient medium or due to intrinsic stellar variability are at the level of $\sim 2.5\%$ only, as given by the fractional variability. Using the theory of two-layer bow-shock \citep{1975ApSS..35..299D}, we are thus only able to place the upper limit on the ambient density at the S2 periapse as well as the upper limit on the density slope. The upper limit considering the contribution of the ambient bow shock, which is more likely to contain dust, is $n_{\rm a}\lesssim 1.87\times 10^9\,{\rm cm^{-3}}$ and the upper limit for the slope is $\gamma\lesssim 3.20$. These values change only slightly when one considers the smaller extinction of ($A=0.02$) as indicated by the excess variance, $n_{\rm a}\sim 7.15 \times 10^8\,{\rm cm^{-3}}$ and $\gamma=3.03$. In case the stellar-wind shock layer would contain dust, which would require the turbulence and the mixing of shocked layers, then the density and the slope constraints decrease to $n_{\rm a}=7.53\times 10^4-1.86 \times 10^5\,{\rm cm^{-3}}$ and $\gamma=1.41-1.57$.  The non-thermal synchrotron emission of the bow-shock appears to put tighter constraints on the S2 pericentre density, $n_{\rm a}\lesssim 1.01\times 10^5\,{\rm cm^{-3}}$ for the maximum flux difference and $n_{\rm a}\lesssim 1.88 \times 10^4\,{\rm cm^{-3}}$ for the fractional variability. For the slope, we obtain $\gamma \lesssim 1.47$ for the maximum flux difference and $\gamma \lesssim 1.17$ for the fractional variability.

Therefore, the $L'$-band data is consistent with the S2 star interacting with the hot, diluted ambient flow with the power-law density slope of $\gamma \sim 0.5$ \citep[generally the Advection Dominated flow - ADAF,][]{2013Sci...341..981W} as well as with the spherical Bondi-type solution with the density slope of $\gamma \sim 1.5$ \citep[also consistent with the Convection Dominated flow - CDAF,][]{2015AA...581A..64R}. Moreover, we cannot exclude the possibility that S2 would interact with an even denser and colder type of the medium, such as the proposed cool disc with densities in the interval of $n_{\rm CD}\sim 10^5-10^6\,{\rm cm^{-3}}$ \citep{2019Natur.570...83M}. In fact, the uppermost limit for the density $n_{\rm a}<1.87\times 10^9\,{\rm cm^{-3}}$ is well in the range for the number density of broad-line region (BLR) clouds $n_{\rm BLR}\sim 10^{8}-10^{11}\,{\rm cm^{-3}}$ \citep{2009NewAR..53..140G}. Although we do not expect the presence of BLR clouds in such a low-luminosity nucleus as Sgr~A* (see, however, \citeauthor{2019MNRAS.488L...1B}, \citeyear{2019MNRAS.488L...1B} for the discovery of the compact BLR in the low-luminosity Seyfert galaxy NGC3147 with $L/L_{\rm Edd}\sim 10^{-4}$), one cannot exclude the presence of denser gaseous-dusty structures, with which stars can occasionally interact,  such as dust-enshrouded objects monitored in the S-cluster \citep{peissker2020,2020Natur.577..337C}. This is also supported by the multiphase medium of Sgr~A* on the scale of one parsec \citep{2017AA...603A..68M}, which could also be present on smaller scales due to e.g., thermal instability \citep{2017MNRAS.464.2090R}.  In particular, \citet{2020arXiv200603648P} report a detection of the proper motion of a Br$\gamma$ filament, whose estimated distance is close to the Bondi radius at $\sim 0.2\,{\rm pc}$. At the comparable scales, \citet{2019ApJ...872....2R} detected a blue-shifted, ionised gas traced using H30$\alpha$ line with the velocity between $-480$ and $-300\,{\rm km\,s^{-1}}$ that appears to be outflowing. Hence, the environment close to the Bondi radius is complex in terms of both kinematics and density.

\begin{figure*}
    \includegraphics[width=\textwidth]{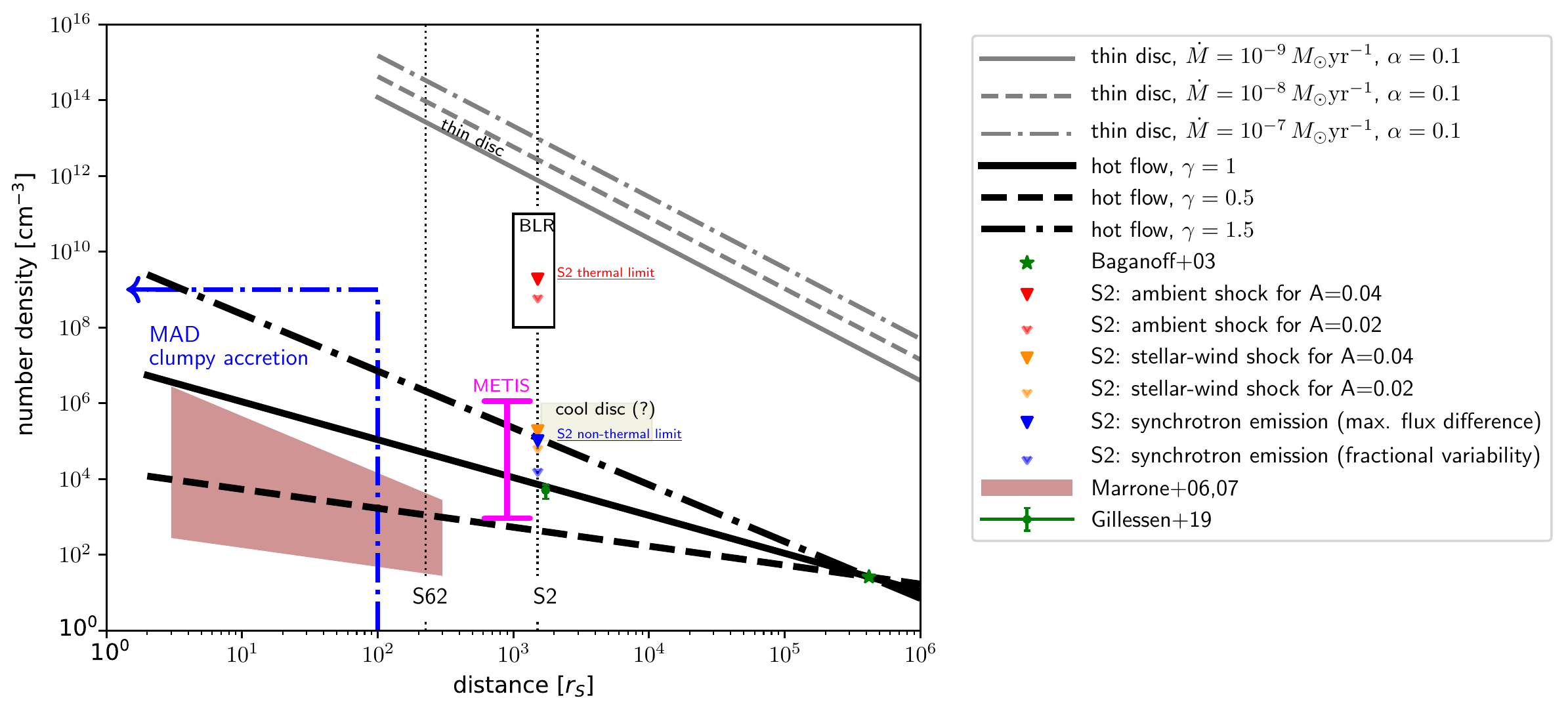}
    \caption{Comparison of the upper density limits that are marked as downward arrows inferred from the $L'$-band light curve of the S2 star based on modelling ambient and stellar-wind shocks for the dust-extinction values of $0.04$ and $0.02$ mag with different density profiles and structures according to the plot legend and the description. Alongside the density limit based on the thermal bow-shock emission, we also depict the limit based on the non-thermal bow-shock emission (bluish downward arrows). For comparison, we show the density limit of $26\,{\rm cm^{-3}}$ inferred by \citet{2003ApJ...591..891B} based on the analysis of the X-ray thermal bremsstrahlung profile. Then we show the density constraint of $(5 \pm 3) \times 10^{3}\,{\rm cm^{-3}}$ inferred from the potential hydrodynamic drag acting on the DSO/G2 object \citep{2019ApJ...871..126G}. The general density power-law profiles are depicted by black lines: the solid line stands for the slope of $1.0$, the dashed line represents the flat profile of $\gamma=0.5$, and the dot-dashed line depicts the steeper Bondi-like spherical flow with $\gamma=1.5$. The shaded rectangle shows the distance and the density limits ($n\sim 10^5-10^6\,{\rm cm^{-3}}$) of the putative cool disc whose discovery is claimed by \citet{2019Natur.570...83M}. The black-framed rectangle depicts the typical densities of the broad-line region (BLR) clouds in the range of $n\sim 10^8-10^{11}\,{\rm cm^{-3}}$. The gray lines depict the number density values expected for a thin disc \citep{1973AA....24..337S} for the accretion rate of $\dot{M}=10^{-9}-10^{-7}\,{\rm M_{\odot}\,yr^{-1}}$ and the viscosity parameter of $\alpha=0.1$. The brown-shaded area to the lower left shows the set of density profiles determined using the rotation measure of the submillimeter polarisation measurements \citep{2006ApJ...640..308M,2007ApJ...654L..57M} for the inner radii of $r_{\rm in}=300\,r_{\rm s}$ and $r_{\rm in}=3\,r_{\rm s}$, density slopes of $\gamma=0.5-1.5$, and the accretion rates in the range $2\times 10^{-9}-2\times 10^{-7}\,M_{\odot}\,{\rm yr^{-1}}$. The blue dot-dashed vertical at 100\,$r_{\rm s}$ and the blue dot-dashed arrow to the left denote the region of a potential magnetically arrested flow (MAD), where the accretion flow changes from continuous to clumpy due to the accumulated poloidal magnetic flux \citep[see e.g.][]{2003PASJ...55L..69N,2019arXiv191208174T}. The dotted vertical lines mark the periapsis of the S62 star at $\sim 225\,r_{\rm s}$ and the periapsis of the S2 star at $\sim 1512\,r_{\rm s}$. Next to the S2 pericentre vertical line, we highlight the range of densities potentially being explored using $L'$-band METIS@ELT imaging mode.}
    \label{fig_density_overview}
\end{figure*}
 Based on the flat $L'$-band light curve and the upper limits on the density, we can exclude the presence of a cold, thin accretion disc \citep{1973AA....24..337S}, whose number densities close to the periapse of S2 at $\sim 1500\,r_{\rm s}$ can be estimated using the relation for the number density as given by the standard accretion disc theory; see also \citet{2002apa..book.....F}. Using the mean atomic weight of $\mu=0.615$ for the fully ionised gas, we obtain,
 
 \begin{align}
     n_{\rm thin} &=2.7\times 10^{12} \left(\frac{\alpha}{0.1} \right)^{-7/10}\left(\frac{\dot{m}}{10^{-8}\,M_{\odot}{\rm yr^{-1}}} \right)^{11/20} \times\notag\\
     &\times \left(\frac{M_{\bullet}}{4\times 10^6\,M_{\odot}} \right)^{5/8} \left(\frac{R}{1500\,r_{\rm s}} \right)^{-15/8}f^{11/5}\,{\rm cm^{-3}}\,,\label{eq_thin_disc}
 \end{align}
 where the factor $f$ stands for $f=[1-(r_{\rm s}/R)^{1/2}]^{1/4}$.
 We calculate the number density range of $n_{\rm disc}=7.6 \times 10^{11}-9.6\times 10^{12}\,{\rm cm^{-3}}$ for the accretion rate between $\dot{m}=10^{-9}-10^{-7}\,M_{\odot}\,{\rm yr^{-1}}$ and the viscous parameter of $\alpha=0.1$ for the S2 pericentre distance. In Fig.~\ref{fig_density_overview}, we plot the thin disc density profiles up to $100\,{\rm r_{\rm s}}$, since the density profile below this radius is rather uncertain and the accretion flow likely breaks up due to the accumulation of poloidal magnetic flux, which results in the magnetically arrested disc \citep[MAD;][]{2003PASJ...55L..69N}. This transition occurs where the magnetic energy density and the gravitational potential energy are at balance, $B^2/8\pi=GM_{\bullet}\rho_{\rm flow}/R_{\rm m}$, from which we can simply derive the magnetospheric radius,
  \begin{align}
    R_{\rm m} & \sim  \frac{8\pi G M_{\bullet} \rho_{\rm flow}}{B^2} =  \\ \,
              & =  90.5 \left(\frac{M_{\bullet}}{4\times 10^6\,M_{\odot}}\right) \left(\frac{n_{\rm flow}}{10^6\,{\rm cm^{-3}}}\right) \left(\frac{B_{\rm pol}}{10\,{\rm G}}\right)^{-2}\, r_{\rm s}\,,\label{eq_magnetospheric_radius}
\end{align}
 where the poloidal magnetic field is scaled to 10 G and the number density of the flow to $10^6\,{\rm cm^{-3}}$ according to the synchrotron models of the Sgr~A* flares \citep{2006ApJ...650..189Y,2012A&A...537A..52E}. The putative MAD part of the flow is highlighted by the blue vertical line in Fig.~\ref{fig_density_overview}.
 
 The thin disc is expected to contain dust from the sublimation radius towards larger distances. If we associate the effective thin disc temperature, $T_{\rm eff}=[3GM_{\bullet}\dot{m}/(8 \pi r^3 \sigma_{\rm B})]^{1/4}$, with the dust sublimation temperature of $T_{\rm sub}=1000\,{\rm K}$, we get a relation for the dust sublimation radius \citep[see also][]{2011AA...525L...8C}
\begin{equation}
    r_{\rm sub}=163 \left(\frac{M_{\bullet}}{4\times 10^6\,M_{\odot}}\right)^{1/3}\left(\frac{\dot{m}}{10^{-7}\,M_{\odot}{\rm yr^{-1}}}\right)^{1/3}\left(\frac{T_{\rm sub}}{1000\,{\rm K}}\right)^{-4/3}\,r_{\rm s}.
    \label{eq_dust_sublimation}
\end{equation}
We obtain the range of $r_{\rm sub}=35-163\,r_{\rm s}$ for the accretion rate range of $\dot{m}=10^{-9}-10^{-7}\,M_{\odot}\,{\rm yr^{-1}}$. Hence, the dusty bow-shock layer should be formed when the S2 would plunge through the disc midplane close to its periapse at $\sim 1500\,r_{\rm s}$. The formation of the shock in the disc material would lead to the extinction increase in $L'$-band magnitude by $0.65$ mag for the density of $4.9 \times 10^{11}\,{\rm cm^{-3}}$ and by as much as $2.30$ mag for the density of $6.13\times 10^{12}\,{\rm cm^{-3}}$. We can therefore exclude the possibility that the S2 star crosses a thin, dense and cold disc.

 Although the comparison with the thin disc density in Fig.~\ref{fig_density_overview} may seem purely academic for the Sgr~A* environment, it is still instructive to compare our upper limits with the thin-disc density for two reasons. First, although the presence of a thin disc was implicitly excluded already based on X-ray and mm observations \citep{2003ApJ...598..301Y,2003ApJ...591..891B,2010ApJ...717.1092D}, the upper density limit inferred from the stellar bow shock is independent from these previous measurements. Second, the transition between the hot advection dominated accretion flows (ADAFs) and thin discs at such low accretion-rate values as for Sgr~A* is still not well understood. For $\alpha$ discs, the viscously stable branch at low $\dot{m}$ exists in the stability curve of surface density--accretion rate or $\Sigma$-$\dot{m}$ plot and these discs would be both geometrically thin and optically thin; see e.g. \citet{2002apa..book.....F}. They would primarily cool via the free-free emission instead of blackbody radiation \citep{1981AcA....31..127T} and one of the main reasons why such a disc is likely not present for Sgr~A* is the short time for plasma to cool down via bremsstrahlung due to the low-angular momentum of the flow. On the other hand, they are low accretion-rate systems where the thin disc was detected, e.g. the maser source NGC4258 \citep{1999ApJ...516..177G} and NGC3147 \citep{2019MNRAS.488L...1B}. Most likely the boundary conditions of the flow, i.e., either feeding by stellar winds or the inflow of gas from larger scales, determine the transition between the ADAF and the thin disc at low $\dot{m}$ and vice versa. In addition, hybrid models of discs have been suggested -- with the outer thin disc part and the inner hot ADAF part, with the transition at 10--100 gravitational radii \citep{1999ApJ...516..177G}.

In order to better constrain the density and lower the upper density limit and the slope, the photometry precision needs to improve below $0.01$ mag. Also, a more precise theory for the bow shock formation and its corresponding NIR emission needs to be taken into account. However, the current methodology that we provide here can be directly transferred to the more precise photometry.

 From future instruments, the Mid-infrared ELT Imager and Spectrograph (METIS) at the Extremely Large Telescope (ELT) will be in particular suited for high-precision photometry in near- and mid-infrared domains. All observing modes of METIS will initially work with a single conjugate adaptive optics (SCAO) system at or close to the diffraction limit of the 39-meter ELT. At a wavelength of 3.5 $\mu$m, the ELT will allow operation at an angular resolution of 23 milliarcseconds. Here, METIS will have a point source sensitivity (10-$\sigma$ in 1 hour) of 21.2 mag (or about 1 $\mu$Jy) \citep{Brandletal2018,2018cwla.conf...41B}. In the N band at a wavelength around 10 $\mu$m, the point-source sensitivity can be expected to be about 1.6 orders of magnitude lower. The field of view of about  10\arcsec $\times$ 10\arcsec\, in imaging and 1.0\arcsec $\times$ 0.5\arcsec\, in IFU (Integral Field Unit) spectroscopy mode at L- and M-bands is ideally suited for the densely populated Galactic Centre region. However, the limiting sensitivity for flux density measurements in this field will be determined by confusion due to source crowding. Hence, the effective sensitivity available to determine flux densities and to establish light curves can safely be assumed to be an order of magnitude lower than the nominal point-source sensitivity, but will still be about an order of magnitude better than the currently reached sensitivity.  For specific estimates of the potentially probed density range, we consider the limiting flux density value for the bow-shock emission in the $L'$-band to be $0.01\,{\rm mJy}$, which corresponds to $\sim 10^{-3}$ mag difference in the S2 star $L'$-band magnitude. Taking into account the model of thermal ambient-shock emission, which is the least sensitive in terms of the density, we obtain the value of $\sim 1.5\times 10^6\,{\rm cm^{-3}}$. This provides an improvement by three orders of magnitude in comparison with the current upper limit. The more sensitive non-thermal bow-shock emission model implies the lower ambient-density limit of only $\sim 700\,{\rm cm^{-3}}$, which is two orders of magnitudes smaller than current limits in Table~\ref{tab_density}. The density range probed by METIS is also shown specifically in the summary plot in Fig.~\ref{fig_density_overview}. In particular, using the lower limit of $700\,{\rm cm^{-3}}$ inferred from the non-thermal model, METIS should be able to directly detect a non-thermal emission increase in case the hot-flow slope is larger than $\gamma=0.5$ between the Bondi radius and the S2 pericentre distance.
Further details will have to be determined through observations on the target.\\

The best next target to observe both the potential thermal and the non-thermal bow-shock emission at the periapse would be S62 in the early 2023 which appears to be a suitable timing to observe the closest S-star to the SMBH by the JWST. It will be an opportunity to shed some more light on the density of the accretion flow around Sgr~A* along the S62 orbital plane \citep{Peissker_S62}, in particular at its pericentre distance of $r_{\rm P}^{S62}\approx 225\,r_{\rm s} \sim 18\,{\rm AU}$, where the conditions of the accretion flow were only constrained via the rotation measure \citep{2006ApJ...640..308M,2007ApJ...654L..57M}, see Fig.~\ref{fig_density_overview}. The detection of the bow-shock emission of S62 would provide an independent means of assessing the ambient number density on the scale of $\sim 100\,r_{\rm s}$. In fact, the S stars S2 and S62 as well as the dusty source DSO/G2 orbit Sgr~A* at different inclinations. Therefore detecting the bow-shock emission for each of them would enable us to constrain the density profile of the accretion flow around Sgr~A* not only as a function of distance, but for a broader range of inclinations.\\

 Near-infrared bow-shock emission from the S stars, which is expected to be most prominent at the pericentre, could be detected via peculiar, longer $K_{\rm s}$ or $L'$-band flares. In fact, the total bow-shock flare duration would be of the order of a few months up to one year and the peak flux is coincident with the pericentre of the star, see our model predictions in Fig.~\ref{fig_light_curve_mag} for the thermal dust emission and in Fig.~\ref{fig_S2_synchrotron} for the non-thermal synchrotron emission associated with the bow shock. The start of the flux increase would be half a year before the pericentre. In terms of the shocked stellar wind, \citet{2013MNRAS.433L..25G} and \citet{2016MNRAS.459.2420C} calculated the timescale of $\sim$ one month for the X-ray bremsstrahlung flare produced by S2 bow-shock with the luminosity of $4\times 10^{33}\,{\rm erg\,s^{-1}}$. However, \citet{Schartmann+2018} argued that for the expected accretion-flow density, the X-ray emission from the bow shock is beyond the detection limit. \citet{Ginsburg+2016} calculated the non-thermal synchrotron emission, which for the typical parameters assumed for S2, the mass-loss rate of $\dot{m}_{\rm w}\sim 10^{-7}\,{\rm M_{\odot} yr^{-1}}$ and the terminal wind velocity of $v_{\rm w}\sim 10^{3}\,{\rm km\,s^{-1}}$, is below the quiescent emission of Sgr~A*. Therefore, the stochastic infrared flares of Sgr~A* \citep{2012ApJS..203...18W,2018ApJ...863...15W}, including the very bright ones \citep{2019ApJ...882L..27D}, are not associated with the stellar bow shocks in the S cluster since they evolve on the timescale of several hours only, which can be interpreted as an orbital timescale of plasmoids close to the innermost stable circular orbit \citep{2018AA...618L..10G}. In other words, some short-term NIR flares could be associated with short-period stars orbiting Sgr~A* on the scale of the innermost stable circular orbit \citep{2020ApJ...896...74L}, which could be disentangled from the stochastic background via their periodic or quasi-periodic signal. However, the bright infrared flare analysed by \citet{2019ApJ...882L..27D} could be associated with the change of an accretion state due to the recent pericentre passage of S2 in 2018 or even the pericentre passage of DSO/G2 in 2014. The time-delay between the pericentre and the average flare statistics may be interpreted via the viscous timescale of the hot accretion flow, which can be of the order of one to ten years \citep{2013AA...555A..97C}.
\section{Conclusions}
\label{summary}
  For the brightest star in the S cluster, S2 star, we obtained a light curve in the near-infrared $L'$-band, which is flat within the measurement uncertainties with the fractional flux variability of only $2.52\%$ in terms of the $L'$-band flux density. When we associate the flux density standard deviation of $0.04$ mag with the local dust extinction due to the bow shock, we can place an upper limit of $n_{\rm a}<1.87 \times 10^9\,{\rm cm^{-3}}$ on the ambient number density at $\sim 1500\,r_{\rm s}$ for the warm bow shock driven into the ambient medium, which is expected to contain dust. For the colder and the denser stellar-wind shock, the number density limit is $n_{\rm a}<1.86 \times 10^5\,{\rm cm^{-3}}$, which is, however, less firm due to the limited presence of dust in stellar winds of B-type stars. Considering the non-thermal bow-shock emission, which does not rely on the existence of dust, we can place an upper density limit of $1.01 \times 10^{5}\,{\rm cm^{-3}}$.

These limits cannot yet constrain the type of the hot accretion flow and hence both the radiatively inefficient accretion flow with the density profile of $\gamma=0.5$ and the spherical Bondi-type flow with $\gamma=1.5$ can accommodate these number densities. We can, however, firmly exclude the standard, thin accretion disc extending all the way to $1\,500\,r_{\rm s}$, whose densities would be at least two orders of magnitude larger and the star-disc interactions would produce a shock with the local extinction of $0.65-2.30$ mag in $L'$ band, which is much larger than the inferred 1$\sigma$ excess of $0.04$ mag of the $L'$-band light curve as well as the difference of $0.11$ mag between the maximum and the minimum magnitude. 

In terms of future prospects, the challenging part is to detect and directly resolve bow shocks associated with the S-stars in near-infrared bands because of the high confusion and stellar density close to Sgr~A*. It may become possible when high-resolution mid-infrared imaging becomes available with the next generation of instruments, namely METIS at the ELT, since bow shocks are generally more prominent due to the dust emission in mid-infrared bands. High-precision photometry associated with METIS with the flux sensitivity of $\sim 0.01$ mJy in $L'$-band is expected to improve the density constraints by two to three orders of magnitude in comparison with our current upper limits. For the non-thermal bow-shock emission, the direct detection of the flux excess is plausible in case the hot-flow density slope is larger than $\gamma\sim 0.5$ between the Bondi radius and the S2 pericentre distance.

\section*{Acknowledgements}
   The authors would like to thank the referee for his/her constructive suggestions and Persis Misquitta for the diligent proofreading of this paper.
  This work was supported in part by SFB 956—Conditions and Impact of Star Formation. S. Elaheh Hosseini is a member of the International Max Planck Research School for Astronomy and Astrophysics at the Universities of Bonn and Cologne. We thank the Collaborative Research Centre 956, sub-project A02, funded by the Deutsche Forschungsgemeinschaft (DFG) – project ID 184018867. Michal Zajacek acknowledges the financial support from the National Science Centre, Poland, grant No. 2017/26/A/ST9/00756 (Maestro 9).

\begin{appendix} 

\section{Flux table}
As described in Section \ref{Photometry results} , we did photometry for the single exposures of the observation 0101.B-0052(B) in 2018.307. As it is shown in Fig. \ref{2018-flux} there is no significant change in flux at the position of S2 and IR counterpart of Sgr~A* therefore, it shows that Sgr~A* is in its quiescent period in $L'$-band during the observation epoch of 2018.307.	
\begin{table}
\caption{Dereddened fluxes of S2 and S65 in $L'$-band in 2018.307}            
\label{table:L_band22}      
\centering          
\begin{tabular}{c c c c c}  
\hline\hline  
UT time &\multicolumn{1}{p{1.5cm}}{\centering S2 \\ (mJy) } & \multicolumn{1}{p{1.5cm}}{\centering S65 \\ (mJy) } \\ 
\hline
05:57:34.4735 	&	9.84	$\pm$	0.11	&	14.48	$\pm$	0.25	\\
05:59:43.8697 	&	9.00	$\pm$	0.02	&	12.38	$\pm$	0.06	\\
06:00:25.8029 	&	10.03   $\pm$	0.27	&	15.61	$\pm$	0.62	\\
06:02:34.9990 	&	1.63	$\pm$	0.05	&	2.94	$\pm$	0.23	\\
06:03:17.5402 	&	8.15	$\pm$	0.01	&	12.43	$\pm$	0.05	\\
06:05:26.7352 	&	8.68	$\pm$	0.04	&	13.34	$\pm$	0.17	\\
06:06:08.8654 	&	9.15	$\pm$	0.13	&	13.81	$\pm$	0.35	\\
06:08:16.4643 	&	10.21   $\pm$	0.38	&	15.21	$\pm$	0.83	\\
06:09:01.9952 	&	9.64	$\pm$	0.26	&	14.70	$\pm$	0.63	\\
06:11:10.5931 	&	8.72	$\pm$	0.10	&	13.86	$\pm$	0.36	\\
06:45:59.1812 	&	10.07   $\pm$	0.13	&	13.38	$\pm$	0.26	\\
06:48:07.7927 	&	9.90	$\pm$	0.10	&	14.93	$\pm$	0.24	\\
06:48:49.5321 	&	12.52   $\pm$	0.63	&	15.67	$\pm$	0.92	\\
06:53:54.6491 	&	10.21   $\pm$	0.04	&	14.05	$\pm$	0.08	\\
06:56:44.5710 	&	10.41   $\pm$	0.18	&	14.11	$\pm$	0.36	\\
06:57:29.3005 	&	8.13	$\pm$	0.03	&	12.32	$\pm$	0.20	\\
06:59:38.1146 	&	8.63	$\pm$	0.03	&	13.97	$\pm$	0.13	\\
07:34:08.2162 	&	8.03	$\pm$	0.05	&	13.4	$\pm$	0.26	\\
07:37:02.3746 	&	8.14	$\pm$	0.03	&	12.95	$\pm$	0.17	\\
07:39:11.7988 	&	8.26	$\pm$	0.02	&	12.55	$\pm$	0.14	\\
07:39:54.9392 	&	8.75	$\pm$	0.06	&	12.57	$\pm$	0.22	\\
07:42:48.1058 	&	7.82	$\pm$	0.17	&	12.75	$\pm$	0.87	\\
07:44:58.3101 	&	8.76	$\pm$	0.06	&	14.46	$\pm$	0.21	\\
07:47:53.4375 	&	8.27	$\pm$	0.17	&	14.10	$\pm$	0.69	\\
08:26:49.3672 	&	7.40	$\pm$	0.03	&	12.60	$\pm$	0.32	\\
08:28:59.1627 	&	7.93	$\pm$	0.03	&	12.82	$\pm$	0.02	\\
08:29:42.3116 	&	9.18	$\pm$	0.06	&	14.00	$\pm$	0.24	\\
08:31:51.9218 	&	8.37	$\pm$	0.02	&	12.53	$\pm$	0.11	\\
08:32:36.0531 	&	8.61	$\pm$	0.08	&	12.88	$\pm$	0.29	\\
08:34:47.4474 	&	10.03   $\pm$	0.13	&	13.62	$\pm$	0.26	\\
08:35:30.1790 	&	8.73	$\pm$	0.11	&	14.57	$\pm$	0.42	\\
08:38:23.3029 	&	8.79	$\pm$	0.25	&	15.14	$\pm$	0.88	\\
08:40:33.6955 	&	7.45	$\pm$	0.03	&	13.72	$\pm$	0.03	\\
09:16:05.5965 	&	8.28	$\pm$	0.03	&	13.75	$\pm$	0.13	\\
09:18:58.3609 	&	7.72	$\pm$	0.02	&	14.01	$\pm$	0.18	\\
09:21:09.3821 	&	7.09	$\pm$	0.03	&	13.30	$\pm$	0.34	\\
09:21:52.5156 	&	7.66	$\pm$	0.02	&	13.09	$\pm$	0.24	\\
09:24:03.3459 	&	7.49	$\pm$	0.02	&	12.31	$\pm$	0.23	\\
09:24:46.0790 	&	9.80	$\pm$	0.25	&	16.09	$\pm$	0.66	\\
09:26:55.2871 	&	8.83	$\pm$	0.05	&	13.62	$\pm$	0.17	\\
09:27:38.2308 	&	8.46	$\pm$	0.10	&	13.89	$\pm$	0.49	\\
09:29:47.4548 	&	8.89	$\pm$	0.07	&	14.16	$\pm$	0.22	\\
10:05:26.0274 	&	8.25	$\pm$	0.02	&	14.11	$\pm$	0.08	\\
10:07:35.2335 	&	8.32	$\pm$	0.10	&	16.33	$\pm$	0.41	\\
10:08:18.1657 	&	8.71	$\pm$	0.10	&	13.24	$\pm$	0.46	\\
10:10:30.5604 	&	9.35	$\pm$	0.18	&	14.43	$\pm$	0.49	\\
10:11:11.8925 	&	9.24	$\pm$	0.03	&	15.21	$\pm$	0.12	\\
10:13:20.2894 	&	8.34	$\pm$	0.03	&	12.99	$\pm$	0.19	\\
10:14:03.2225 	&	8.48	$\pm$	0.02	&	13.07	$\pm$	0.12	\\
10:16:13.2227 	&	8.44	$\pm$	0.09	&	12.47	$\pm$	0.47	\\
10:19:10.3576 	&	10.80   $\pm$	0.47	&	16.81	$\pm$	1.12	\\                
\hline                
\end{tabular}
\end{table}
\section{The S2 star bow-shock evolution along the orbit: formula for the density slope }
\label{derivation}
The size of a bow shock formed due to the supersonic motion of the star close the supermassive black hole is scaled by the standoff distance $R_0$ that depends on two stellar parameters --- the mass-loss rate $\dot{m}_{\rm{w}}$ and the terminal wind velocity $v_{\rm{w}}$ --- and the density of the ambient medium $\rho_{\rm{a}}$ as well as the relative velocity of the star with respect to the surrounding medium $v_{\rm{rel}}$. In the most general form, it can be expressed in the following way \citep{1996ApJ...459L..31W,1997ApJ...474..719Z,2016MNRAS.459.2420C},
\begin{equation}
  R_0=\left(\frac{\dot{m}_{\rm{w}} v_{\rm{w}}}{\Omega (1+\alpha) \rho_{\rm{a}} v_{\rm{rel}}^2}\right)^{1/2} \simeq C_{\star}v_{\rm{rel}}^{-1}\rho_{\rm{a}}^{-1/2}\,,
  \label{eq_stagnation_radius}
\end{equation}
where $\Omega$ stands for the solid angle into which the stellar wind is blown (in case it is fully isotropic $\Omega=4\pi$). The coefficient $\alpha$ is the ratio of thermal and the ram pressure of the ambient medium, $\alpha=P_{\rm{th}}/P_{\rm{ram}}$. The supersonic motion takes place when the Mach number is larger than one, $M=v_{\rm{rel}}/c_{\rm{s}}=1/\sqrt{\kappa \alpha}>1$, where $c_{\rm{s}}$ is the sound speed and $\kappa$ is an adiabatic index. The last equality in Eq.~\ref{eq_stagnation_radius} is valid for the supersonic limit when $\alpha \rightarrow 0$, i.e. when the thermal pressure is low.  The term $C_{\star}=(\dot{m}_{\rm{w}} v_{\rm{w}}/\Omega)^{1/2}$ then represents a stellar parameter that we assume to be constant during one orbital period, which for S-stars with a semi-major axis $a$ typically is (in years),
\begin{equation}
  \tau_{\rm{orb}}=11.85\,\left(\frac{a}{0.1''} \right)^{3/2} \left(\frac{M_{\bullet}}{4\times 10^6\,M_{\odot}} \right)^{-1/2}\,\text{yr}\,.
  \label{eq_period}
\end{equation}

\begin{figure}[tbh]
  \centering
  \includegraphics[width=0.5\textwidth]{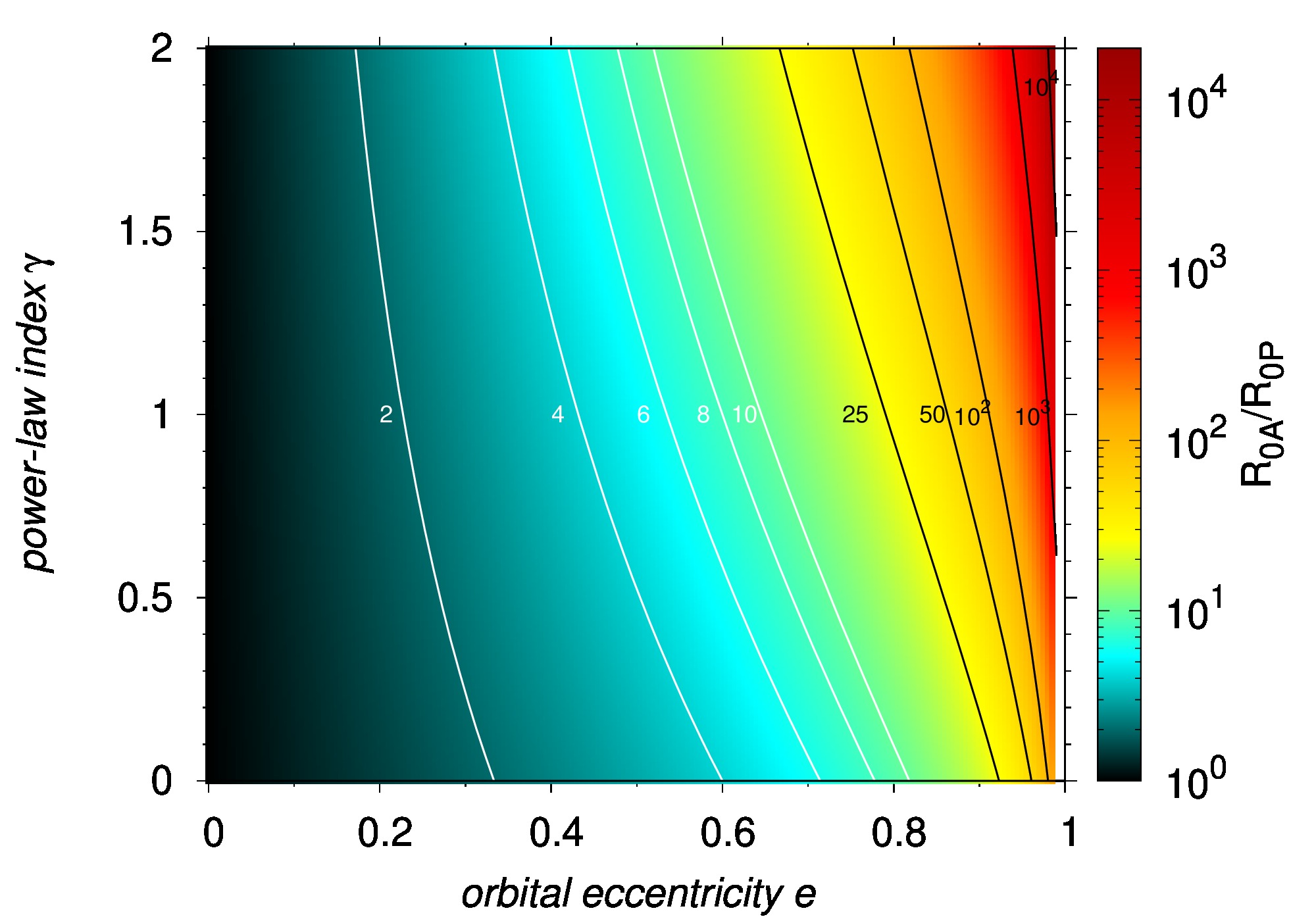}
  \caption{The colour-coded ratio $R_{0A}/R_{0P}$ of stellar bow-shock sizes between the apocentre and the pericentre of the corresponding orbit in the $e$-$\gamma$ plane (see also Eq.~\ref{eq_ratio_final}). The white and black lines correspond to the contours of the constant ratio with numbers indicating the value of $R_{0A}/R_{0P}$ for a given contour.}
  \label{fig_gamma_e}
\end{figure}

The stellar parameter $C_{\star}$ may be evaluated for the mass-loss rate $\dot{m}_{\rm{w}}$ and the terminal wind velocity $v_{\rm{w}}$ as inferred for S2 star based on spectroscopic information \citep{2008ApJ...672L.119M},
\begin{equation}
  C_{\star}=7.1\times 10^{12}\left(\frac{\dot{m}_{{\rm w}}}{10^{-7}\,M_{\odot}\,{\rm yr}^{-1}}\right)^{1/2} \left(\frac{v_{{\rm w}}}{10^3\,{\rm km\,s^{-1}}}\right)^{1/2}\,{\rm g^{1/2}\,cm^{1/2}\,s^{-1}}.
  \label{eq_stellar_parameter}   
\end{equation}
Now we consider the relation between the scale-lengths of the bow shock $R_{0}$ at two different positions. We assume that the number density profile near the Galactic Centre as a function of distance $r$ from the black hole has a power-law form:
\begin{equation}
  n_{\rm{a}}=n_{0}\left( \frac{r}{r_0} \right)^{-\gamma}\,,
  \label{eq_powerlaw}
\end{equation}
where the power-law index $\gamma$ is positive, $\gamma \geq 0$. It follows directly from Eq.~\eqref{eq_stagnation_radius} that the ratio of the standoff radii between two positions of the bow shock along the orbit is as follows,
\begin{equation}
  \frac{R_{01}}{R_{02}}=\frac{v_{\rm{rel2}}}{v_{\rm{rel1}}}\left(\frac{\rho_{\rm{a2}}}{\rho_{\rm{a1}}}\right)^{1/2}=\frac{v_{\rm{rel2}}}{v_{\rm{rel1}}}\left(\frac{n_{\rm{a2}}}{n_{\rm{a1}}}\right)^{1/2}=\frac{v_{\rm{rel2}}}{v_{\rm{rel1}}}\left(\frac{r_{1}}{r_{2}}\right)^{\gamma/2}\,,
  \label{eq_ratio_R01_R02}
\end{equation}
where the subscripts $1$ and $2$ stand for two positions along the elliptical orbit characterised by velocities $v_{\rm{rel1}}$ and $v_{\rm{rel2}}$ and distances from the focus (SMBH) $r_1$ and $r_2$ $(r_1>r_2)$, respectively. The last equality in Eq.~\eqref{eq_ratio_R01_R02} is derived using a general expression for the power-law density profile, Eq.~\eqref{eq_powerlaw}.
 Unless otherwise indicated in the text, we consider the ambient medium to be stationary with respect to the star, therefore the relative velocity is approximately given by the orbital velocity of the star around the Galactic centre, $v_{\rm{rel}}\simeq v_{\star}$. Hence, we obtain a slightly modified relation for the ratio,
\begin{equation}
  \frac{R_{01}}{R_{02}}=\frac{v_{\star 2}}{v_{\star 1}}\left(\frac{n_{\rm{a2}}}{n_{\rm{a1}}}\right)^{1/2}=\left(\frac{r_1}{r_2}\right)^{(\gamma+1)/2}\left(\frac{2a-r_2}{2a-r_1}\right)^{1/2}\,.
  \label{eq_ratio1}
\end{equation} 
The last equality is obtained using the expression for the orbital velocity of a star on an elliptic orbit, or the vis-viva equation, $v_{\star}=\sqrt{\mu(2/r-1/a)}$, where $\mu=GM_{\bullet}$.
In case the orbital elements for a given S star are constrained, one can invert Eq.~\ref{eq_ratio1} to get the power-law index $\gamma$ as function of the bow-shock standoff ratio at two positions $R_{01}(r_1)/R_{02}(r_2)$, orbital distances $r_1$ and $r_2$, and the semi-major axis $a$,
\begin{equation}
 \gamma_{12}=\frac{2 \log{[R_{01}(r_1)/R_{02}(r_2)]}+\log{[(2a-r_1)/(2a-r_2)]}}{\log{(r_1/r_2)}}-1\,. 
 \label{eq_power_law_index}
\end{equation}
The number of parameters in relation \ref{eq_power_law_index} may be reduced when evaluated at special points on the orbit, especially the pericentre (true anomaly $\nu=0^{\circ}$), the apocentre ($\nu=180^{\circ}$), and the semi-latus rectum ($\nu=90^{\circ}$). From the classical celestial mechanics, we easily obtain the ratios for corresponding distances and velocities at the apocentre and the pericentre for a general elliptical orbit with an eccentricity $e$. The ratio of the distances from the supermassive black hole is $(r_{\rm{P}}/r_{\rm{A}})=(1-e)/(1+e)$\footnote{We denote the quantities evaluated at the pericentre, apocentre, and the semi-latus rectum by indices $P$, $A$, and $90^{\circ}$, respectively.} and the ratio of velocities is $(v_{\rm{orbP}}/v_{\rm{orbA}})=(1+e)/(1-e)$. 
Thus, the ratio between the apocentre and the pericentre bow-shock size is simply a function of two parameters --- the orbital eccentricity $e$ and the power-law index for the ambient density distribution $\gamma$,
\begin{equation}
   \frac{R_{0A}}{R_{0P}}=\left(\frac{1+e}{1-e}\right)^{1+\gamma/2}\,.
   \label{eq_ratio_final}
\end{equation}
In the same way as in Eq.~\ref{eq_power_law_index}, one can express the power-law index in the following way,
\begin{equation}
  \gamma_{AP}=2\big[\log{(R_{0A}/R_{0P})}/\log{[(1+e)/(1-e)]}-1\big]\,.
  \label{eq_gamma_apo_peri}
\end{equation}
A similar simple expression is obtained when evaluating the ratio between the points at the latus-rectum intersection with the ellipse and the pericentre. We denote the standoff distance at that point $R_{0,90}$ (the true anomaly is 90 degrees). The following relation then holds,
\begin{equation}
   \frac{R_{0,90}}{R_{0P}}=\frac{(1+e)^{(\gamma+2)/2}}{(1+e^2)^{1/2}}\,,
   \label{eq_ratio_latus}
\end{equation}  
from which the density power-law index trivially follows (or from Eq.~\eqref{eq_power_law_index}),
\begin{equation}
  \gamma_{LP}=\frac{2 \log{[R_{0,90}/R_{0P}]}+\log{(1+e^2)}}{\log{(1+e)}}-2\,.
  \label{eq_gamma_latus_peri_plus}
\end{equation}
Since the star passes through the medium with a variable density $n_{\rm{a}}$, the bow-shock emission is also expected to change. An upper limit for the thermal bow-shock emission is given by the rate of thermalized kinetic terms of the stellar and ambient winds, $L_{\rm{th}} = 1/2 \dot{m}_{\rm{w}} (v_{\rm{rel}}^2+v_{\rm{w}}^2) \simeq 1/2 \dot{m}_{\rm{w}} (v_{\star}^2+v_{\rm{w}}^2)$ \citep{1997IAUS..182..343W}. The ratio of thermal bow-shock luminosities between two points is then given by,
\begin{equation}
 L_{\rm{th 1}}/ L_{\rm{th 2}}=(\beta_1^2+1)/(\beta_2^2+1)\,,
 \label{ratio_lum}
\end{equation}
where $\beta=v_{\star}/v_{\rm{w}}$.
Table~\ref{table_ratio} summarises different exemplary values of the index $\gamma$ and the corresponding values of the ratio of bow-shock sizes between the apocentre and the pericentre -- Eq.~\ref{eq_ratio_final} -- as well as the ratio of luminosities of thermal emission, Eq.~\ref{ratio_lum}, for different eccentricities. As a standard prototype of S stars, we take S2 with inferred $a\simeq 0.123''$, $e\simeq0.88$, and $v_{\rm{w}}=1000\,\rm{km\,s^{-1}}$ \citep{2008ApJ...672L.119M,2017ApJ...847..120H,2018AA...615L..15G}.
\begin{table}[tbh]
  \centering
  \begin{tabular}{cccc}
     \hline
     \hline
     $e$ & $\gamma$ & $(R_{0A}/R_{0P})_0$ (Eq.~\eqref{eq_ratio_final}) & $L_{\rm{thA}}/L_{\rm{thP}}$ (S2)\\
     \hline
      $0$      &      $\geq 0$ &       $1$     &       $1$  \\
      $0.5$    &      $0.0$    &      $3.0$    &       $0.19$\\     
      $0.5$    &      $0.5$    &      $3.9$    &       $0.19$\\     
      $0.5$    &      $1.0$    &      $5.2$    &       $0.19$\\     
      $0.5$    &      $2.0$    &      $9.0$    &       $0.19$\\     
      $0.9$    &      $0.0$    &      $19.0$   &       $1.8 \times 10^{-2}$\\ 
      $0.9$    &      $0.5$    &      $39.7$   &       $1.8 \times 10^{-2}$\\
      $0.9$    &      $1.0$     &     $82.8$   &       $1.8 \times 10^{-2}$\\
      $0.9$    &      $2.0$    &      $361.0$  &       $1.8 \times 10^{-2}$\\
    \hline  
  \end{tabular}
  \caption{The power-law index $\gamma$ and the corresponding ratio between the apocentre and the pericentre bow-shock sizes and the thermal emission luminosities calculated for different eccentricities.}
  \label{table_ratio}
\end{table}
In Fig.~\ref{fig_gamma_e}, we plot the colour-coded ratio $R_{0A}/R_{0P}$ in the $e$-$\gamma$ plane (eccentricity--power-law index). We see that in general the ratio $R_{0A}/R_{0P}$ has a broad range between $1$ up to $10^{4}$ for plausible values of the density power-law slope, $\gamma \in (0,2)$, and the orbital eccentricity, $e\in (0,0.99)$. For the mean eccentricity of $e=2/3$ in the thermal (isotropic) distribution of eccentricities, $n(e)\mathrm{d}e=2e\mathrm{d}e$, and the flat density profile of $\gamma=0.5-1.0$ \citep{2013Sci...341..981W}, the ratio $R_{0A}/R_{0P}$ is between $7.5$ and $11.2$ according to Eq.~\eqref{eq_ratio_final}.
\textit{Effect of thermal pressure.} In the analysis above, we considered the thermal pressure of the ambient medium to be negligible with respect to the ram pressure, i.e. $\alpha=P_{\rm{th}}/P_{\rm{ram}} \rightarrow 0$. Yet, at different distances from the SMBH, the factor $\alpha(r)=P_{\rm{th}}/P_{\rm{ram}}=c_{\rm{s}}(r)^2/v_{\star}(r)^2$ may in principle change and become non-negligible. The ratio of standoff radii between the apocentre and the pericentre of the orbit then becomes,
\begin{equation}
  \frac{R_{0A}}{R_{0P}}=\left(\frac{R_{0A}}{R_{0P}}\right)_0 \left(\frac{1+\alpha_{\rm{P}}}{1+\alpha_{\rm{A}}}\right)^{1/2}\,,
   \label{eq_ratio_therm1}
\end{equation}  
where $(R_{0A}/R_{0P})_0$ is the ratio with a neglected thermal pressure, Eq.~\ref{eq_ratio_final}. The thermal term at the pericentre may be evaluated as follows,
\begin{equation}
 \alpha_{\rm{P}}=c_{\rm{s}}(r_{\rm{P}})^2/v_{\star}(r_{\rm{P}})^2=\frac{\kappa k_{\rm{B}} T_{\rm{P}}}{\overline{\mu} m_{\rm{p}}} \frac{a}{\mu} \frac{1-e}{1+e}\,
 \label{eq_therm_peri}
\end{equation}
where $T_{\rm{P}}$ is the gas temperature at the pericentre, $\overline{\mu}$ is the mean molecular weight, and $m_{\rm{p}}$ is a proton mass. The ratio of the thermal terms at the apocentre and the pericentre then is,
\begin{equation}
  \frac{\alpha_{\rm{A}}}{\alpha_{\rm{P}}}=\frac{T_{\rm{A}}}{T_{\rm{P}}}\left(\frac{1+e}{1-e}\right)^2=\left(\frac{r_{\rm{P}}}{r_{\rm{A}}}\right)^{\delta}\left(\frac{1+e}{1-e}\right)^2=\left(\frac{1+e}{1-e}\right)^{2-\delta}\,,
  \label{eq_ratio_ram_pressures}
\end{equation}
where $\delta \geq 0$ is a power-law index of the spherical temperature profile of the ambient medium, $T_{\rm{a}}(r)=T_0(r_0/r)^{\delta}$, where $T_0=T_{\rm{a}}(r_0)$. The extra term in Eq.~\eqref{eq_ratio_therm1} would vanish if $\alpha_{P}=\alpha_{A}$, which would occur if  $e=0$. Using Eq.~\eqref{eq_ratio_ram_pressures} we may rewrite Eq.~\eqref{eq_ratio_therm1} into the form,
\begin{equation}
  \frac{R_{0A}}{R_{0P}}=\left(\frac{R_{0A}}{R_{0P}}\right)_0 \underbrace{\left[\frac{1+\alpha_{\rm{P}}(T_{\rm{P}},a,e)}{1+\alpha_{\rm{P}}(T_{\rm{P}},a,e)\left(\frac{1+e}{1-e}\right)^{2-\delta}}\right]^{1/2}}_{F_{AP}(T_{\rm{P}},a,e,\delta)}\,,
   \label{eq_ratio_therm}
\end{equation} 
For the estimation of the factor $F_{AP}$ in brackets that arises due to thermal pressure, we consider the orbital elements of S2 star and the following temperature profile of the ambient medium \citep{2012ApJ...759..130P,2011ApJ...735..110B},
\begin{equation}
  T_{\rm{P}}=T_0 \left(\frac{GM_{\bullet}}{rc^2}\right)^{\delta}\,,
  \label{eq_temp_prof}
\end{equation}
where $\delta=0.84$ and $T_0=9.5 \times 10^{10}\,\rm{K}$. After calculation, the correction term is $\simeq 0.8$ for S2 star, so every ratio in Table~\ref{table_ratio} should be slightly adjusted, $R_{0A}/R_{0P} \simeq 0.8 (R_{0A}/R_{0P})_0$, if the temperature profile, Eq.~\eqref{eq_temp_prof}, holds. 
Including the thermal pressure term, the formula for the density power-law index $\gamma$ based on the ratio of the apocentre and the pericentre standoff radii changes accordingly,
\begin{equation}
  \gamma'_{AP}=2\big[(\log{(R_{0A}/R_{0P})}-\log{F_{AP}})/\log{[(1+e)/(1-e)]}-1\big]\,.
  \label{eq_gamma_apo_peri_therm}
\end{equation}
Similarly, using the semi-latus rectum--pericentre ratio $R_{0,90}/R_{0P}$, one can derive the slope $\gamma'_{LP}$ modified by the thermal therm $F_{LP}$ in the following way,
\begin{equation}
  \gamma'_{LP}=\frac{2 [\log{(R_{0,90}/R_{0P})}-\log{F_{LP}}]+\log{(1+e^2)}}{\log{(1+e)}}-2\,,
  \label{eq_gamma_latus_peri_minus}
\end{equation}
where the thermal term is expressed by
\begin{equation}
 F_{LP}(T_{{\rm p}},a,e)=\left[\frac{1+\alpha_{{\rm P}}(T_{{\rm P}},a,e)}{1+\alpha_{{\rm P}}(T_{{\rm P}},a,e)\frac{(1+e)^{2-\delta}}{(1-e)^2}}\right]^{1/2}\,.
\end{equation}
\end{appendix}
\end{document}